\documentclass{nature}
\usepackage[utf8]{inputenc}
\usepackage{amsmath}
\usepackage{amsfonts}
\usepackage[nocomma]{optidef}
\usepackage[euler]{textgreek}
\usepackage{multirow}
\usepackage{float}
\usepackage{tfrupee}
\usepackage{gensymb}
\floatstyle{plaintop}
\DeclareUnicodeCharacter{2212}{-}
\restylefloat{table}

\usepackage[labelfont=bf]{caption}
\makeatother

\title{Sustained cost declines in solar PV and battery storage needed to eliminate coal generation in India}

\author{Aniruddh Mohan$^{1*}$, Shayak Sengupta$^{1}$, Parth Vaishnav$^{1,2}$, Rahul Tongia$^{1,3}$, Asim Ahmed$^{4}$, Ines L. Azevedo$^{5}$}

\usepackage{graphicx}
\makeatletter
\let\saved@includegraphics\includegraphics
\AtBeginDocument{\let\includegraphics\saved@includegraphics}

\begin{document}

\maketitle
\begin{small}
\begin{affiliations}
 \item Department of Engineering and Public Policy, Carnegie Mellon University, Pittsburgh, 15213, USA
 \item School for Environment and Sustainability, University of Michigan, Ann Arbor, 48109, USA
 \item Centre for Social and Economic Progress, New Delhi, 110021, India 
 \item Reconnect Energy, Bangalore, 560008, India
 \item Department of Energy Resources Engineering, Stanford University, Stanford, 94305, USA
\end{affiliations}
\end{small}

\begin{abstract}
Unabated coal power in India  must be phased out by mid-century to achieve global climate targets under the Paris Agreement. Here we estimate the costs of hybrid power plants - lithium-ion battery storage with wind and solar PV - to replace coal generation. We design least cost mixes of these technologies to supply stylized baseload and load-following generation profiles in three Indian states - Karnataka, Gujarat, and Tamil Nadu. Our analysis shows that availability of low cost capital, solar PV capital costs of at least \$250/kW, and battery storage capacity costs at least 50\% cheaper than current levels will be required to phase out existing coal power plants. Phaseout by 2040 requires a 6\% annual decline in the cost of hybrid systems over the next two decades. We find that replacing coal generation with hybrid systems 99\% of the hours over multiple decades is roughly 40\% cheaper than 100\% replacement, indicating a key role for other low cost grid flexibility mechanisms to help hasten coal phaseout. Solar PV is more suited to pairing with short duration storage than wind power. Overall, our results describe the challenging technological and policy advances needed to achieve the temperature goals of the Paris Agreement.
\end{abstract}

More ambitious emission reduction targets for greenhouse gases (GHG) than those submitted by countries' Nationally Determined Contributions (NDCs) are needed to achieve the temperature goals of the Paris Agreement \cite{masson2018global, du2018warming, van2017early, peters2017key}. This will require accelerated deployment of clean energy solutions such as wind power, solar power, and energy storage technology \cite{mccollum2018energy, peters2017key}.

Previous modelling studies on the future of India's electricity sector have considered various scenarios with high levels of renewable energy penetration \cite{deshmukh2021least, palchak2017greening, gulagi2017electricity}. Such studies are usually based on capacity expansion and dispatch models that meet energy demand with investments in different energy sources. The complexity of these models inevitably requires several assumptions such as a well-functioning national electricity market and limited transmission, political, or institutional constraints. The relevance of such approaches is however limited in developing country contexts such as India, where electricity is not sold through fully deregulated markets, and where political and institutional constraints can be barriers to technology adoption \cite{iychettira2021lessons}. 

Instead, our study focuses solely on a pivotal climate policy issue relevant for India's emissions trajectory and global emissions pathways - the possibilities of reducing coal power use in India through development of alternate energy sources that can provide the same generation profile to the electricity grid. We make minimal assumptions around the rest of the electricity system in India apart from the assumption of continued requirement of baseload and/or flexible power generation sources. We build an optimization model to study the least cost combinations of energy storage with wind and solar power that can provide such power generation profiles. Our study, the first of its kind for India, builds on previous work studying the value and operation of energy storage systems for decarbonizing the U.S. electricity system \cite{ziegler2019storage, braff2016value}. 

We estimate that the levelized costs of systems that provide flexible or baseload generation every hour over a twenty year period are currently in the range of \rupee 10-14/kWh, assuming a lower cost of capital. To be competitive with currently operating or new coal power plants, costs must fall by at least 60\%. An annual cost decline of 6\% in both solar PV and battery storage could enable phase out of coal power beginning 2040 in India, broadly consistent with the power sector decarbonization goals necessary to limit global average temperature rise to 1.5\degree C \cite{cui2019quantifying, iea}. This pace of cost decline can also avoid the construction of new coal power plants in the 2030s.

\section*{Results}
We focus on three states in India: Karnataka, Gujarat, and Tamil Nadu. These three states have relatively higher coal generation costs compared to other states as they are located further from coal mines which increases coal procurement costs\cite{tongia2019coal, kamboj2018indian}. They are also considered part of the better performing group of Indian states in terms of renewable energy deployment\cite{busby2021solar, thapar2018key} with installed renewable energy capacity that is in the top five of all Indian states\cite{mnre2021}. We draw on twenty years of hourly wind and solar output for site locations in these three states. Table \ref{table1} shows the average capacity factor (CF) for wind energy and solar energy at the selected site locations in the three Indian states we consider. For more details on the wind and solar resource data see Methods.

\begin{table}[ht]
  \begin{center}
    \caption{\textbf{Average CF for wind and solar at our selected site locations in Karnataka, Rajasthan, and Gujarat}}\label{table1}
    \vspace{0.5cm}
    \begin{tabular}{l|l|r} 
      \textbf{State} & \textbf{Technology} & \textbf{Average CF}\\
      \hline
      \multirow{2}{*}{Karnataka} & 
      Wind  & 22\% \\
      & Solar  & 19\% \\
    \hline
      \multirow{2}{*}{Tamil Nadu} & 
      Wind  &  19\%\\
      & Solar  &  20\%\\
    \hline
      \multirow{2}{*}{Gujarat} & 
      Wind  &  23\%\\
      & Solar  & 20\% \\
    \hline
    \end{tabular}
  \end{center}
\end{table}

\begin{figure}[!h]
    \centering
    \includegraphics[width=13cm, height = 6cm]{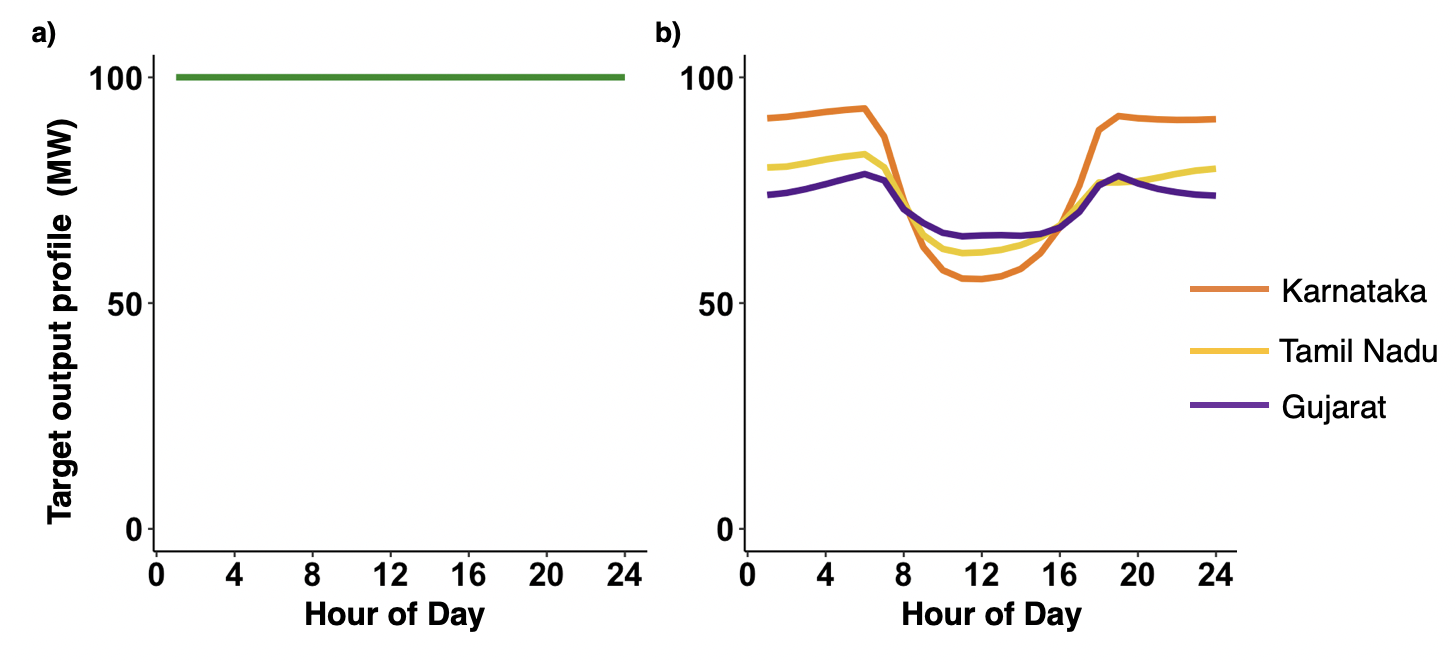}
    \caption{\small \textbf{(a)} Simulated target generation as inflexible baseload operating at 100 MW every hour of the year. \textbf{(b)} Average 24h period of simulated flexible target generation of coal power plants in the three states. Details regarding the flexible target generation profile for each state are provided in Methods.}\label{fig1}
    \vspace{0.5cm}
\end{figure}

We consider two types of stylized generation profiles for coal power as useful bounding cases for estimating the cost of replacing coal power generation. First, the target coal plant to be replaced is assumed to run as inflexible baseload, providing 100 MW of power every hour of the year, as shown in Figure \ref{fig1}(a). While inflexible baseload operation may be a useful limiting case to understand the costs of such systems, coal plants in India are increasingly run in a flexible manner to incorporate higher shares of wind and solar generation in India. At times of high wind and solar production coal generators often back down to lower levels of output, rising again to maximum output during the evening hours as demand rises and solar generation drops off. If the objective is to replace a dispatchable source of power such as flexible coal plants rather than firm inflexible baseload, grid operators may require hybrid power producers to similarly reduce output during hours where standalone wind or solar plants without battery storage are producing at full capacity. We therefore model a second stylized operational profile for coal power shown in Figure \ref{fig1}(b), where coal generation accommodates the variable renewable energy already present in the respective state grids. Details regarding the simulation of the stylized flexible generation profile for each state are provided in Methods. Note that while Figure \ref{fig1} only shows an average 24h period, our optimization model uses the full time varying twenty year hourly target profile, which differs from the 24h average profile in the case of flexible generation.

\begin{figure}[!h]
    \centering
    \includegraphics[width=16cm, height = 8.8cm]{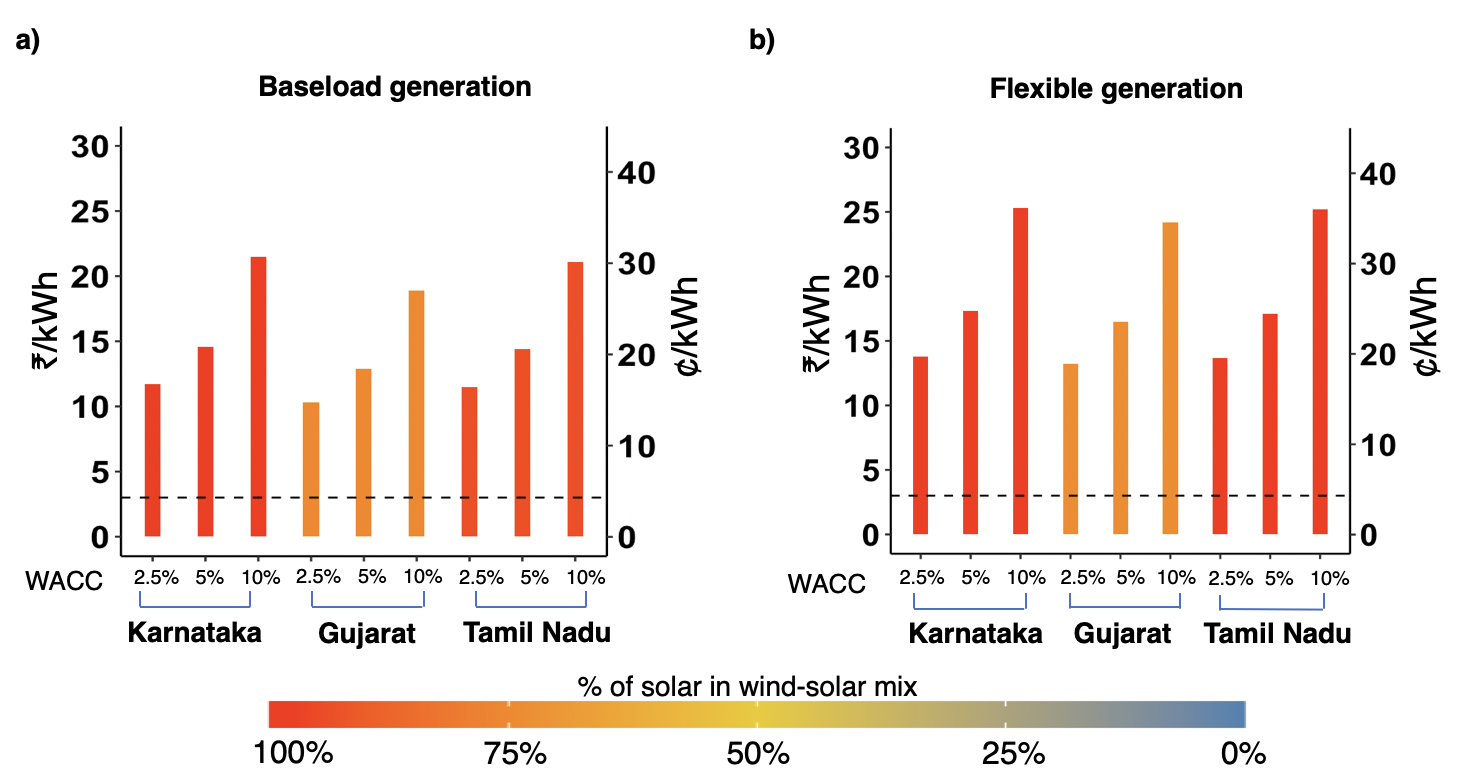}
    \caption{\small Figure 2 shows the LCOE for least cost hybrid systems in Karnataka, Tamil Nadu and Gujarat, for different assumptions around the weighted average cost of capital. LCOE values are shown in both \rupee/kWh and \$/kWh. We assume a fixed conversion rate where \$1=\rupee 70. \textbf{(a)}The LCOE of hybrid systems with an optimal mix of wind power, solar PV and battery storage for each state and WACC that can provide baseload generation shown in Fig 1(a). The colour of the bars indicates the share of solar in the solar-wind optimal mix. The dashed black line represents the marginal cost of coal power plants operating in those states today. \textbf{(b)} The LCOE of hybrid systems with an optimal mix of wind power, solar PV and battery storage for each state and WACC that can provide the flexible coal generation shown in Fig 1 (b). The colour of the bars indicates the share of solar in the solar-wind optimal mix. The dashed black line represents the marginal cost of coal power plants operating in those states today.}\label{fig2}
    \vspace{0.5cm}
\end{figure}

The levelized cost of electricity (LCOE) of a hybrid system with an optimal mix of wind and solar capacity with short duration (4h) lithium-ion storage to provide the respective profiles of generation are shown in Figure \ref{fig2}(a) and Figure \ref{fig2}(b). We estimate the LCOE for different assumptions about the weighted average cost of capital (WACC). Solar, wind and battery storage system capital costs are assumed as \$700/kW, \$1100/kW and \$400/kWh respectively\cite{taylor2020irena, mongird20202020, cole2020cost}. The individual capacities of wind, solar PV, and battery storage that make up the optimal hybrid power plant for each state and type of generation profile are shown in Table \ref{table2}. The results presented in Figure \ref{fig2} and Table \ref{table2} show that solar is dominant in the hybrid power plant for each state and type of generation considered. Solar is found to be more suitable for pairing with lithium-ion battery storage in the states considered given its more predictable output across seasons as well as its lower cost compared to wind. More details on solar and wind generation and robustness checks on our renewable energy data are provided in Methods and SI Supplementary Notes 3-4. 

The results shown in Figure \ref{fig2} indicate that hybrid systems are currently more expensive than the marginal cost (which includes fuel, operations, and maintenance cost) of the coal power plants operating in those states today, represented by the dashed black line at \rupee 3.0/kWh \cite{shrimali2020making, nazar2021implication}. Further, levelized costs of the system are roughly similar across both flexible and inflexible generation profiles. Baseload systems are cheaper as they sell more power to the grid across the operational lifetime and therefore lead to greater utilization of the capital investments. Assuming a lower cost of capital of 2.5\%, baseload and flexible hybrid systems have levelized costs between \rupee 10-14/kWh which is several times the cost of existing coal generation in those states. 

\begin{table}[ht]
  \begin{center}
    \caption{\textbf{Capacity sizes of the hybrid power plants that mix the optimal combination of wind power, solar PV, and 4-hour duration Li-ion battery storage to provide the target profile of generation at least cost.}}\label{table2}
    \vspace{0.5cm}
    \begin{tabular}{l|l|r|r|r} 
      \textbf{State} & \textbf{Generation Type} & \textbf{Wind (MW)} & \textbf{Solar PV (MW)} & \textbf{Battery Storage (MWh)} \\
      \hline
      \multirow{2}{*}{Karnataka} & 
      Baseload  & 14 & 1685 & 2727\\
      & Flexible  & 0 & 1582 & 2553\\
      \hline
      \multirow{2}{*}{Gujarat} & 
        Baseload  & 371 & 1136 & 2038\\
      & Flexible  & 376 & 1058 & 1761\\
     \hline
      \multirow{2}{*}{Tamil Nadu} & 
        Baseload  & 109 & 1550 & 2606\\
      & Flexible  & 0 & 1154 & 2898\\
    \hline
    \end{tabular}
  \end{center}
\end{table}

\subsection{Mixing wind and solar power:}
\begin{figure}[!h]
    \centering
    \vspace{-0.2cm}
    \includegraphics[width=16cm, height = 5cm]{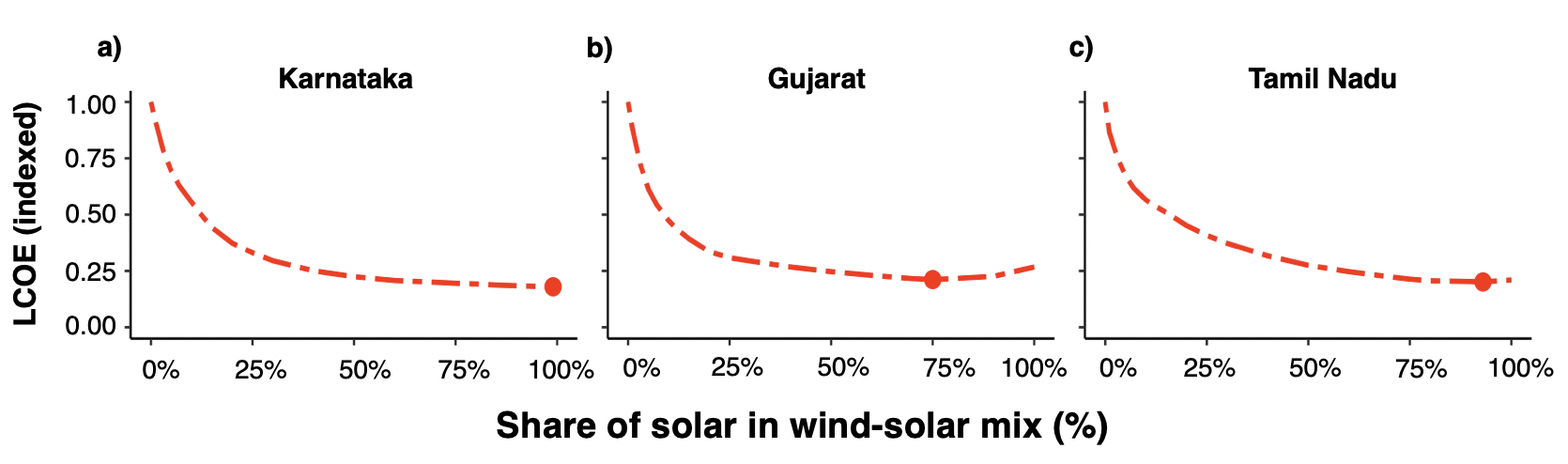}
    \caption{\small Figure 3 shows the LCOE for different shares of solar PV in hybrid systems with both wind and solar PV providing baseload generation. LCOE values on the Y axis are indexed to the LCOE for a hybrid system with only wind power and battery storage. \textbf{(a)} Results for Karnataka. The optimal mix shown with the red dot has 99\% solar PV and 1\% wind capacity. \textbf{(b)} Results for Gujarat. The optimal mix shown with the red dot has 75\% solar PV and 25\% wind capacity. \textbf{(c)} Results for Tamil Nadu. The optimal mix shown with the red dot has 93\% solar PV and 7\% wind capacity.}
    \label{fig3}
    \vspace{0.5cm}
\end{figure}

To understand the relative value of each  intermittent renewable technology, we model different mixes of wind and solar PV in a hybrid system with 4h battery storage. Figure \ref{fig3} shows the change in the levelized cost of the system providing baseload generation for different shares of wind and solar PV capacity in the total renewable energy capacity of the plant.

Costs of hybrid systems which only incorporate wind energy with storage are significantly higher at the sites considered. This is due to the high seasonality of wind generation in India (also see SI Notes 3-4), with high output during monsoon season and very limited output the rest of the year. As such, wind is a poor companion to short duration storage for providing baseload or flexible generation every hour of the year. As solar is added to the hybrid mix, costs fall. However, as Figure \ref{fig3} shows, hybrid systems which combine even small amounts of wind capacity are cheaper than a system with only solar PV and storage for each of the three sites considered. The equivalent analysis for flexible generation is shown in SI Supplementary Note 1 (Figure S1). 

\subsection{Daily operation of battery storage:}
\begin{figure}[!h]
    \centering
    \vspace{0.2cm}
    \includegraphics[width=14.6cm, height = 7.3cm]{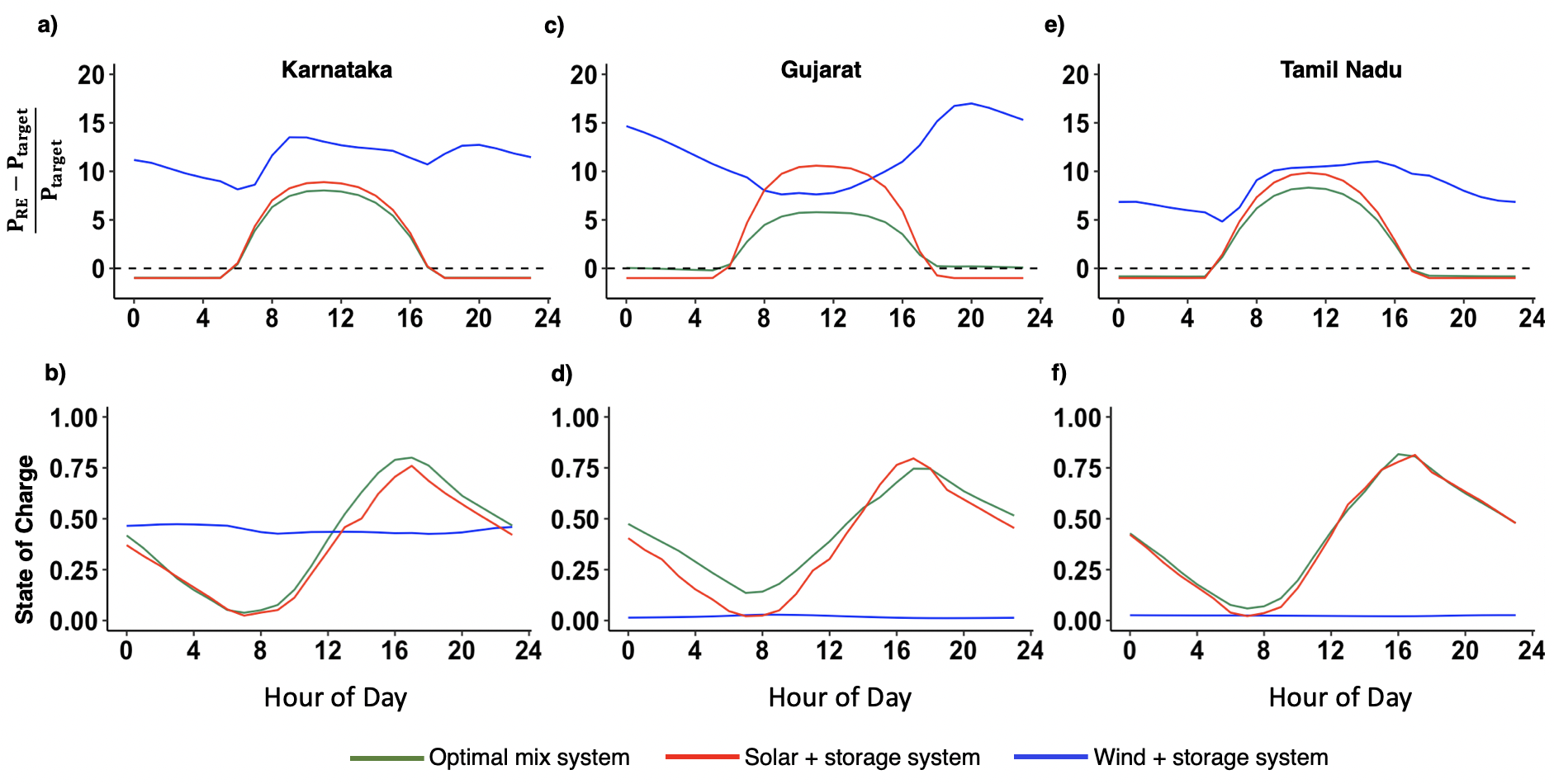}
    \caption{\small Average daily operation of the hybrid plant that provides the target baseload 100 MW generation every hour for 20 years for the state of Karnataka, Gujarat and Tamil Nadu. Significant excess generation from renewable energy allows charging of the battery which discharges to meet the generation requirement when output from wind and solar is low. \textbf{(a,c,e)} The average hourly normalized difference between the production of wind and solar power from the hybrid plant and the target 100 MW output through the day. \textbf{(b,d,f)} The average hourly SOC (\%) of the battery storage system through the day.}
    \label{fig4}
    \vspace{0.25cm}
\end{figure}

Figure \ref{fig4} shows the average 24h operation of the hybrid system including battery storage for providing the baseload target generation profile.

Due to the high share of solar power in the optimal mix in each of the three states, the behaviour of the hybrid system is similar in both the case of an optimal mix system and in a system with solar only paired with storage. This is particularly true for both Karnataka and Tamil Nadu where solar is more than 90\% of the variable renewable energy capacity in the hybrid system. In these cases, the storage system charges through the morning and afternoon hours as solar comes online (Fig \ref{fig4}) leading to a peak in the state of charge (SOC) around 4-5pm in the evening (Fig \ref{fig4}). The system then discharges again through the night and early morning hours (Fig \ref{fig4}). The behaviour of these systems is driven by the shortfall between the production of renewable energy by the wind and/or solar plants through the 24h period and the target hourly 100 MW output as shown in Figure \ref{fig4}(a). As expected, in the case of the solar only system shown by the red line in the Figure, there is maximum excess power generated just around noon when solar power is at maximum. On the other hand, a wind only plant has a flatter production profile and as a result the operation of the battery storage system is also much flatter with less variation in power and SOC through the day (Fig \ref{fig4}). Wind only plants with storage are also significantly oversized due to the high seasonal nature of wind in India with very little wind output for large parts of the year and extremely high output during the monsoon season. As a result, daily averages skew the picture in the case of wind only plants with storage. The full twenty year hourly state of charge of the battery storage system across all three states and all three cases (optimal mix, solar + storage, and wind + storage) is shown in the SI Figures S2-S4 where we also describe the individual capacities of wind, solar, and batteries for each of these cases in the three states. 

The average 24h operation of the battery storage system in Figure \ref{fig4} as well as the optimal capacity sizes for the solar PV plant and wind power plant shown in Table \ref{table2} reflect the basic operational dynamics of hybrid power systems. Solar PV and wind capacities are significantly oversized compared to the target generation required to be met. This enables excess energy during sunny and windy hours to charge the battery. The battery is then discharged during times of low renewable energy production to provide the target generation. However due to the fact that charging and discharging the battery involves a cost in the form of efficiency losses as well as capital costs in terms of larger battery sizing, not all the excess power generated during peak wind and solar production hours is used to charge the battery. We find that across the three states, for the optimal hybrid system replacing baseload or flexible generation, total curtailment is 150-300\% of total required annual generation. That is, even after using excess wind and solar generation to charge the battery storage system, significant amounts of renewable generation is curtailed. Limitless curtailment proves to be a cost effective solution as in a system which instead minimizes curtailment, the LCOE is an order of magnitude higher due to much larger battery sizing. 

\subsection{Reducing the cost of hybrid systems:}
In this section we explore three different levers for reducing the cost of hybrid systems. These include: monetizing excess generation that is normally curtailed; reducing the availability requirement of hybrid plants; and finally drawing on wind and solar resources from a wider region to smooth variation in output.

\subsubsection*{1. Value of curtailment}
The LCOE results presented in Figure \ref{fig2} are based on no economic value being assigned to curtailed energy. However, it may be possible that independent power producers (IPPs) who own and operate such hybrid facilities will be able to sell excess intermittent renewable generation to the grid through separate contracts. For example, grid operators may purchase the excess power treating it as must run intermittent renewable generation. If IPPs are able to contract sale of this excess power generated at even small shares of the sale value of the baseload or flexible generation, this would reduce the overall system LCOE. 

We therefore consider the case where some economic value is assigned to curtailed energy from the oversized hybrid system. Figure S5 in the Supplementary Information shows the change in the LCOE as some of this excess generation is monetized. This excess generation could be valued in different ways. One could value the entire excess generation at a fraction of the target generation produced by the hybrid system. With increasing value assigned, the LCOE drops. Alternatively, excess generation could be assigned the same value as target generation but only some share of it could be sold. Or there could be a combination of both. The Y axis in Figure S5 shows the LCOE for different valuations of the excess generation as a fraction of the target generation shown on the X axis. We find that if 100\% of the power that was curtailed was monetized and valued the same as the target generation, levelized costs that were presented in Figure \ref{fig2} could drop by as much as 70\%. Flexible demand sources in the energy system such as hydrogen electrolysis or electric vehicle charging could be potential buyers of such excess generation, and help lower the costs of hybrid systems.

\subsubsection*{2. Reduced Availability}
Hybrid systems will also be allowed downtime, similar to coal power plants in India that often operate at 60-80\% capacity factor for the year\cite{mohan2019india}. While the results presented in Figure \ref{fig2} are for 100\% availability systems, here we model the levelized costs of hybrid plants with lower than 100\% availability, where the constraint of meeting the required generation profile is relaxed for some hours\cite{ziegler2019storage} in the 20 year period. A 100\% system meets the baseload generation required for all 175200 hours in 20 years. A 90\% system would meet this requirement for 157680 hours. Results for these different scenarios are shown in Figure \ref{fig5} below.

\begin{figure}[!h]
    \vspace{1cm}
    \centering
    \vspace{-0.5cm}
    \includegraphics[width=16cm, height = 6.7cm]{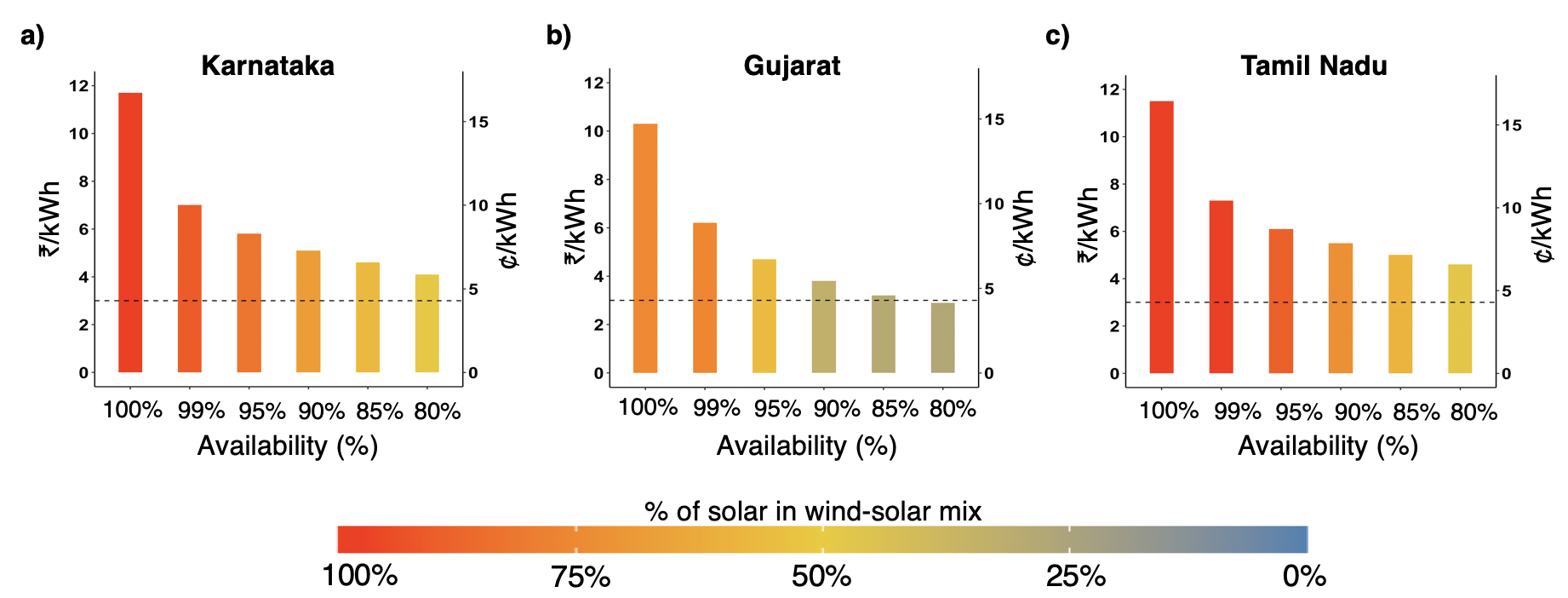}
    \caption{\small{Figure 5 shows the LCOE assuming a 2.5\% cost of capital for different levels of the system availability which is defined as the share of hours over 20 years where the hybrid plant meets the required 100 MW baseload generation. \textbf{a)} Karnataka. \textbf{b)} Gujarat. \textbf{c)} Tamil Nadu. The dashed horizontal black line is the marginal cost of coal power plants operating in those states today.}}
    \label{fig5}
\end{figure}

As the required availability of the system is reduced, levelized costs fall. Crucially, lowering the availability required by just 1\% from 100\% to 99\% leads to a roughly 40\% reduction in levelized costs across the three states. Hybrid systems that can provide baseload generation 99\% of the hours over 20 years are therefore notably cheaper than those that provide the generation 100\% of the hours. This shows that a small share of hours in the 20 year period are driving the high cost of hybrid systems that meet the generation requirement 100\% of the time. Tables S1-S3 in the SI Note 1 show why this is the case. In the 99\% system, the size of the battery storage system is much lower, reducing costs compared to the 100\% system. A small share of hours with prolonged low output from wind and solar over the 20 year period requires bigger battery sizing for the 100\% system, and therefore leads to much higher costs overall. By reducing the availability requirement even slightly, the hybrid system can rely more on wind and solar resources instead of the battery, reducing costs. Costs fall further as the availability required is further lowered as shown in Figure \ref{fig5}. We also find that as the availability required is lowered, the share of wind in the wind-solar mix increases (Figure 5, Tables S1-S3). For example in Karnataka, the hybrid system providing baseload generation in 100\% of hours has just 1\% of wind in the wind-solar mix. However, a 80\% system is a wind dominated hybrid plant with 51\% wind capacity in the wind-solar mix. Similarly for Gujarat, the share of wind goes from 25\% in the 100\% system to 62\% in the 80\% case, with steady increases in the wind share for each level of lower availability. In fact as the availability required is lowered, wind substitutes for storage in the hybrid system, which is what drives the cost reductions. We find that in Gujarat, an optimal mix hybrid system that provides baseload 100 MW power in 80\% of the hours over 20 years has a lower levelized cost than \rupee 3.0/kWh (Figure \ref{fig5}), the target for replacing existing coal generation. We caution however that these results do not necessarily mean that hybrid plants at 80\% capacity factor are cheaper than 80\% capacity factor coal power plants. The issue is that the precise timing of their availability will influence system and therefore consumer costs for grid electricity. The results shown in Figure \ref{fig5} are based on modelling downtime when it optimal for the hybrid plant operator, which due to the relative costs of wind, solar and storage, is when renewable generation is lowest and the battery would be needed most to fulfil the constraint of providing the target 100MW generation. Of course, it is precisely when renewable generation on the electricity grid drops to low levels that coal generation today would likely be running at high output. Therefore, it is unlikely that hybrid plants will be allowed to have downtime in those hours if they are replacing coal power on the grid, as intended here. If so, other resources such as diesel generation or demand response may be required which incur their own costs. As such, we caution that while levelized costs for such systems fall in all the three states as availability is lowered to more realistic operational availability of less than 100\%, this does not necessarily imply that coal generation can be replaced at those lower costs. Nevertheless, the results shown here can shed some light on the influence of system availability on levelized costs and also highlight the opportunities for other low cost grid flexibility mechanisms such as demand response which could potentially cover the 1\% or 10\% of hours over twenty years at a lower cost than the cost difference between 100\% availability and 99\% or 90\% availability hybrid systems.

\subsubsection*{3. Geographical smoothing}
Our results so far have focused on hybrid power plants at a single site, drawing on co-located ($<$100km apart) wind and solar resources in a single location along with a battery storage system. However, wind resources in particular may have high spatial heterogeneity. Drawing from resources in multiple locations could lead to reduced costs, as dips in output in one location could be compensated by another, smoothing the output profile of the variable renewable resource with potentially lower dependence on the battery and an increased role for wind energy.

To model this, we draw on twenty years of hourly wind and solar data in four individual sites in the northern, southern, eastern, and western parts of each state. The exact locations for each of the states are shown in SI Figure S6. We model the levelized costs of hybrid power plants at each of these individual site locations. Then, we consider a fifth case, by averaging the hourly wind and solar output across these four individual locations to what could potentially represent a state averaged renewable resource, and model a hybrid system based on this state averaged resource. Comparisons between the costs of hybrid systems which can draw output from multiple locations versus the costs of hybrid systems that are located only in individual locations can help shed some light on how geographical smoothing of resources influence the design and costs of hybrid power plants for providing baseload or flexible generation.

Table \ref{table3} below shows the results for the state of Gujarat. Similar results for  Karnataka and Tamil Nadu are shown in Table S4 and S5 in the SI. 

\begin{table}[ht]
\begin{center}
\caption{\textbf{\small Levelized costs and capacity mix of hybrid power systems providing 100 MW baseload power over 20 years in the state of Gujarat at different individual site locations and a state averaged location which averages out resources from the four individual locations. All levelized costs shown are based on a 2.5\% cost of capital. Note that similar results for Karnataka and Tamil Nadu are shown in the SI Table S4 and S5.}}
\begin{tabular}{l|r|r|r|r|r|r|r}
\label{table3}

\textbf{Location} & \textbf{Solar CF} & \textbf{Wind CF} & \textbf{\begin{tabular}[c]{@{}l@{}}Solar \\ (MW)\end{tabular}} & \textbf{\begin{tabular}[c]{@{}l@{}}Wind \\ (MW)\end{tabular}} & \textbf{\begin{tabular}[c]{@{}l@{}}Battery \\ (MWh)\end{tabular}} & \textbf{\begin{tabular}[c]{@{}l@{}}Solar \% in \\ least cost mix\end{tabular}} & \textbf{\begin{tabular}[c]{@{}l@{}}LCOE \\ (Rs/kWh)\end{tabular}} \\
\hline
North      & 25\%     & 27\%    & 1907                                                  & 69                                                   & 2102                                                     & 97\%                                                                  & 11.5                                                     \\
\hline
East       & 24\%     & 27\%    & 827                                                   & 287                                                  & 3489                                                     & 74\%                                                                  & 11.7                                                     \\
\hline
South      & 23\%     & 33\%    & 1617                                                  & 170                                                  & 2351                                                     & 90\%                                                                  & 11.5                                                     \\
\hline
West       & 24\%     & 33\%    & 1271                                                  & 205                                                  & 2114                                                     & 86\%                                                                  & 10.0                                                     \\
\hline
State avg. & 24\%     & 30\%    & 1232                                                  & 138                                                  & 1581                                                     & 90\%                                                                  & 8.4                                                     
\end{tabular}
 \end{center}
\end{table}

We find that smoothing resources over a wider region in Gujarat leads to lower levelized costs, and savings of 16\% over the lowest cost individual location (West) and savings of 28\% over the most expensive individual location considered (East). We caution however that the levelized costs for hybrid power plants based on the state averaged resource do not necessarily fully represent the costs of replacing coal generation. This is because to this cost one must also add the cost of transmission, functioning real time electricity markets, and other soft and hard infrastructure required to integrate renewable resources from different corners of a state. We also note that even with state averaged resources, wind continues to have a minor role, with the least cost hybrid system solar dominated in each of the three states (Table \ref{table3}, SI Table S4,S5). Finally, the costs of such hybrid plants based on state averaged resources nevertheless continue to far exceed the \rupee 3.0/kWh cost target required to compete with existing coal generation.

\subsection{Cost of CO$_{2}$ abatement:}
Avoiding coal generation for the 20 year operational lifetime of the hybrid power plant will avoid several million tonnes of CO$_{2}$ emissions. We calculate the \$/tCO$_{2}$ abated. We assume a 0.92tCO$_{2}$/MWh emissions factor for coal in India \cite{deshmukh2021least} and \rupee 3.0/kWh or \$43/MWh for existing coal generation in India \cite{shrimali2020making, nazar2021implication}. Note that for coal generation, almost all the CO$_{2}$ emissions are associated with combustion with a minor share from upstream processes\cite{venkatesh2012uncertainty}. The reverse is true for battery storage and renewable energy where CO$_{2}$ emissions at the point of generation of electricity are zero but upstream emissions involved in manufacturing and mining of materials are non-zero. We draw on the literature on lifecycle CO$_{2}$ emissions of battery storage, solar PV, and wind energy\cite{sharma2020towards, pehl2017understanding, pucker2021greenhouse} to estimate the emissions factor for our hybrid systems as approximately 0.02tCO$_{2}$/MWh. Using these data we estimate the marginal cost of abatement, which is equal to the difference in generation cost of hybrid power plants and coal power ($\Delta$\$/MWh) divided by the difference in the emissions factor ($\Delta$tCO$_{2}$/MWh). Table \ref{table4} shows the marginal cost of abatement from hybrid power systems for providing different generation profiles and across different cost of capital assumptions for the three states.

\begin{table}[ht]
  \begin{center}
    \caption{\textbf{\$/tCO$_{2}$ avoided from the hybrid system for different assumptions for the weighted average cost of capital and target generation profile}}
    \vspace{0.5cm}
    \label{table4}
    \begin{tabular}{l|l|r|r|r} 
      \textbf{State} & \textbf{Generation} & \textbf{2.5\% WACC} & \textbf{5\% WACC} & \textbf{10\% WACC}\\
      \hline
      \multirow{2}{*}{Karnataka} & 
      Baseload  & 138 & 184 & 293\\
      & Flexible & 171 &  226 &  355 \\
    \hline
      \multirow{2}{*}{Tamil Nadu} & 
       Baseload  & 135 & 180 & 287\\
      & Flexible  &  170 &  224 & 351\\
    \hline
      \multirow{2}{*}{Gujarat} & 
       Baseload  & 116 & 157 & 253\\
      & Flexible  &  162 &  214 & 337\\
    \end{tabular}
  \end{center}
\end{table}

A comparison of these costs with estimates of the social cost of carbon (SCC) can be useful to understand whether such technologies are cost effective solutions to mitigate CO$_{2}$ emissions. There is high uncertainty with regards to the SCC and estimates range many orders of magnitude\cite{wang2019estimates,moore2015temperature} from as low as \$10/tCO$_{2}$ to $>$\$1000 tCO$_{2}$. Perhaps more pertinently, Ricke et al\cite{ricke2018country} provide estimates for country specific marginal damage costs of CO$_{2}$ emissions and find a central estimate of \$86/tCO$_{2}$ in India. Again the estimate spans a wide uncertainty range from \$49/tCO$_{2}$ to \$157/tCO$_{2}$\cite{ricke2018country}. The Interagency Working Group (IWG) in the United States proposes a central estimate of \$51/tCO$_{2}$ in 2020 for regulatory purposes which corresponds to a 3\% rate of discounting future damages\cite{scc2021technical}. Table \ref{table4} shows that the current mitigation costs for hybrid systems across different assumptions for the cost of capital exceed commonly used estimates of the SCC. 

We also undertake an analysis for the benefits of avoided premature mortality from coal generation and this is provided in SI Supplementary Note 2. We find that the benefits are of the order of \rupee 1/kWh. As per the results in Figure \ref{fig2}, this is clearly insufficient to bridge the gap between the current costs of hybrid systems and operating coal power plants.

\subsection{Implications for coal plants and global climate targets\\} 
Most global emission pathways compatible with limiting global average temperature rise to 2\degree C call for complete phaseout of unabated coal power plants by 2050 \cite{cui2019quantifying, johnson2015stranded} while in order to limit temperature rise to 1.5\degree C, unabated coal power generation should end by 2040 \cite{cui2019quantifying,iea}.

\begin{figure}[!h]
    \vspace{1cm}
    \centering
    \vspace{-0.5cm}
    \includegraphics[width=15cm, height = 8cm]{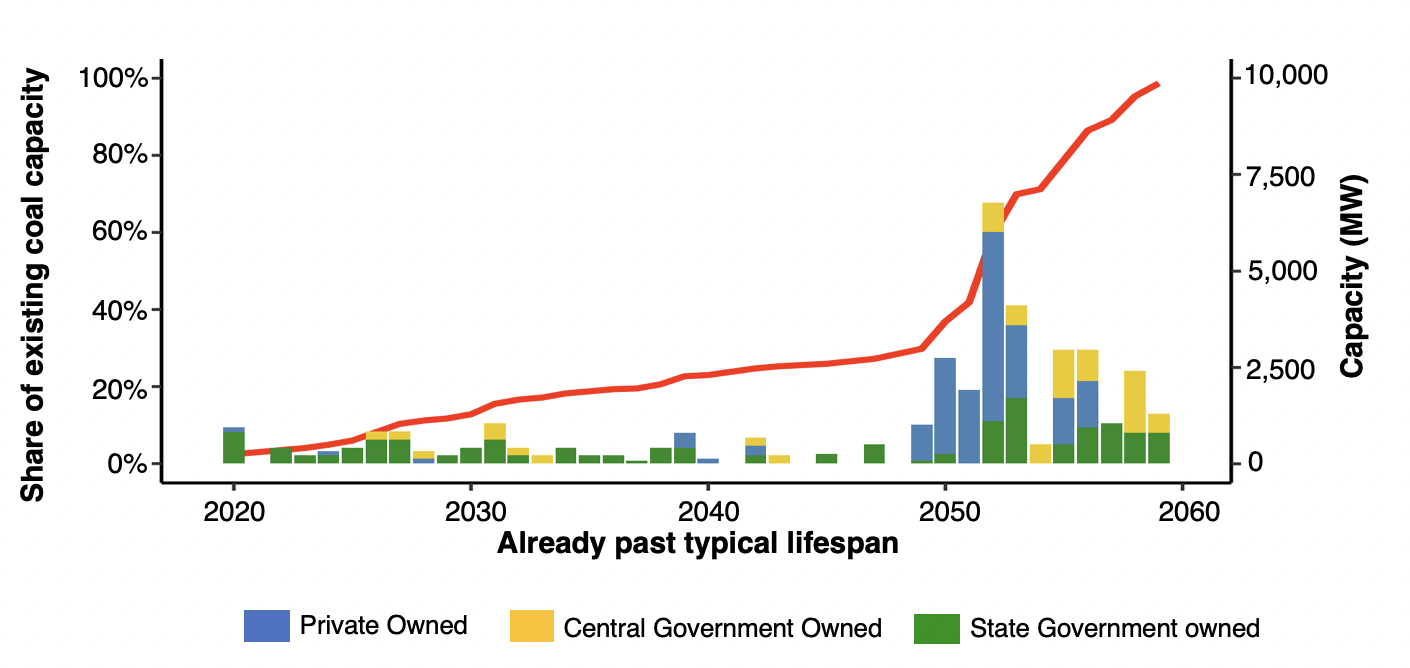}
    \caption{\small Retirement timeline for coal power plants across the three states of Karnataka, Gujarat and Tamil Nadu assuming a natural 40 year plant lifetime. The red line is measured on the Y axis on the left and plots the cumulative share of total capacity that is beyond its typical lifetime. The stacked bars are measured on the secondary Y axis on the right and plot the amount of coal capacity, based on ownership, that is scheduled to retire in each year. Total operational coal capacity across the three states is 39,000 MW. }
    \label{fig6}
    \vspace{0.5cm}
\end{figure}

We assemble a database of currently operational coal power plants in the three states (see Methods). India's coal fleet is relatively young\cite{cui2019quantifying, yang2019future} as are the coal plants in our selected states of Karnataka, Tamil Nadu and Gujarat. We find a capacity weighted average age of just 13 years for the fleet across these states. Assuming a 40 year lifetime for operational plants\cite{yang2019future, shearer2017future}, Figure \ref{fig6} shows the retirement timeline for currently operating coal power plants in Karnataka, Tamil Nadu and Gujarat. 

Figure \ref{fig6} shows that a phaseout of existing plants by both 2040 or 2050 would either require policy intervention or displacement by lower cost alternate sources of generation, given the relatively small share of capacity likely to naturally retire by those dates. Any such phaseout would therefore strand a substantial share of existing coal capacity. This stands in contrast to other countries such as the United States for example, where a majority of coal power capacity is scheduled to naturally retire by 2035\cite{grubert2020fossil}. A 2040 phaseout target in India strands more than 75\% of coal capacity across the three states while a 2050 target leads to earlier than planned retirement for up to 60\% of the fleet. Note that while our analysis shown in Figure \ref{fig6} focuses on plants in three states, high shares of stranded capacity from a 2040 or 2050 coal phaseout in India holds true across the country\cite{cui2019quantifying}. This suggests that implementing policies that can achieve such phaseout targets might be politically and economically unpalatable, particularly given the strong governmental support coal has historically enjoyed in India\cite{mohan2018india}. Naturally, phaseout may prove politically easier if alternate clean technologies such as hybrid systems can provide power at lower costs than existing coal generation, i.e. $<$ \rupee 3.0/kWh \cite{nazar2021implication, shrimali2020making}. From the results shown in Figure \ref{fig2}, that will require significant cost reductions from today's costs across the three states, even with assuming a lower cost of capital of 2.5\%.


\subsection{Future cost reductions\\}
Wind, solar PV, and lithium-ion battery storage have all seen rapid cost declines in recent years\cite{taylor2020irena,ziegler2021re} which provide evidence towards future lower costs for hybrid systems. For instance, global weighted installation costs of wind and solar PV have fallen 24\% and 79\% respectively between 2010-2019 \cite{taylor2020irena}. Meanwhile, lithium-ion battery pack prices have fallen 89\% between 2010-2020 \cite{bnef}.

We examine the future cost targets these technologies  (wind, solar PV, battery storage) would have to hit, to enable levelized generation costs that can economically displace coal generation and achieve the goals of the Paris Agreement. To understand various scenarios of displacement, we consider different cost scenarios for coal generation. Recent governmental directives have called for the introduction of pollution control technologies (PCTs) that could increase marginal costs by up to 25\% \cite{nazar2021implication}. Current cost of coal generation in India also does not account for the health or climate externalities imposed by coal power, aside from a small \rupee 400/ton cess on coal production. We therefore consider cases where a carbon tax as low as \$20/tCO$_{2}$ to as high as \$60/tCO$_{2}$ is imposed, decreasing the relative competitiveness of coal versus alternate generation technologies. For the levelized cost of new coal generation we assume costs of roughly \rupee 4/kWh \cite{tongia2019coal}. We consider baseload hybrid systems with 99\% availability, i.e. hybrid systems that provide baseload 100 MW power for 99\% of the hours over a twenty year period. We focus here on 99\% and not 100\% systems as it clearly conceivable that in the future other low cost grid flexibility mechanisms could handle the lean 1\% of hours, which as shown previously in Figure \ref{fig5} significantly increase the cost of hybrid power plants.

\begin{figure}[!h]
    \vspace{1cm}
    \centering
    \vspace{-0.5cm}
    \includegraphics[width=17cm, height = 10cm]{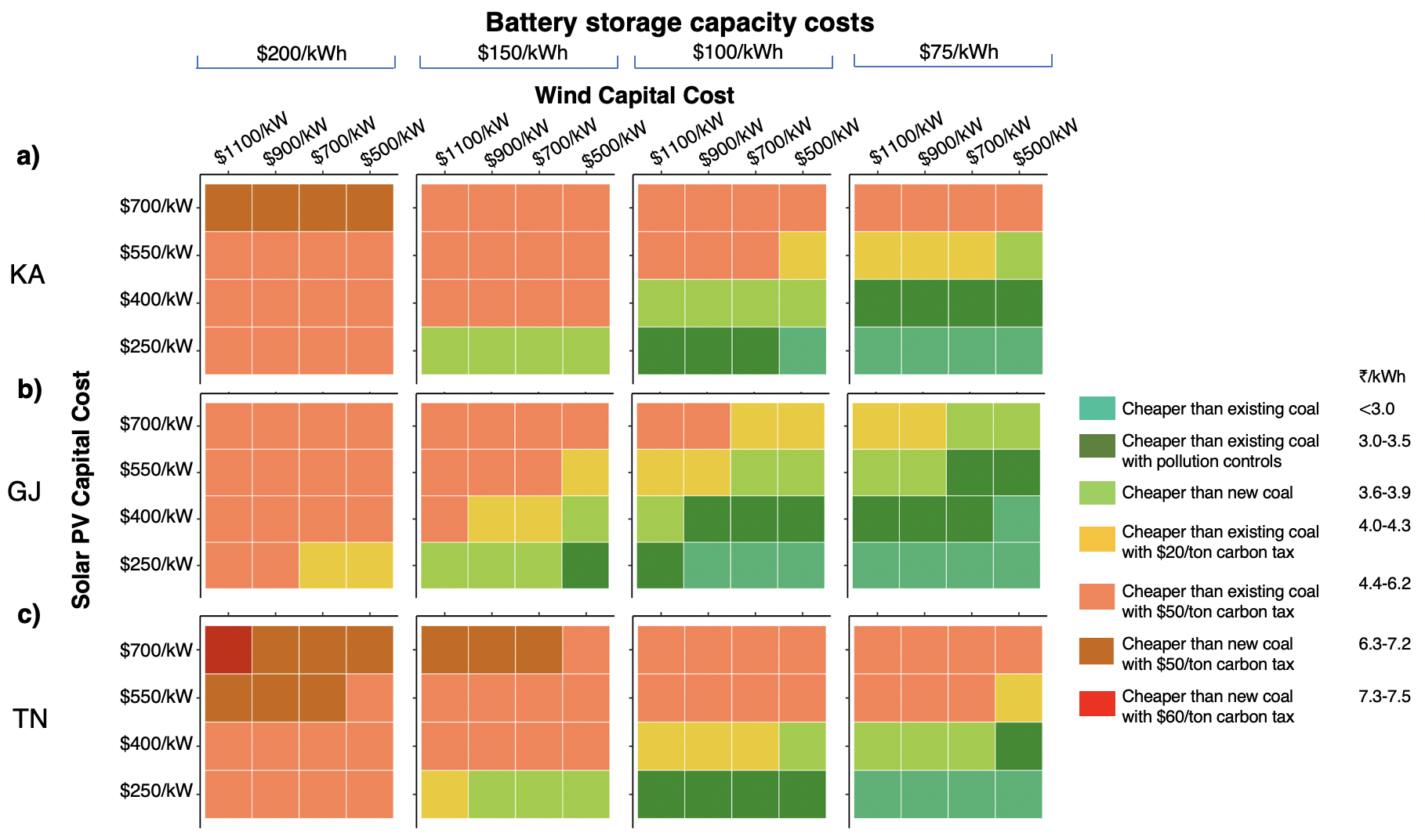}
    \caption{\small Figure shows the levelized costs of hybrid systems that provide baseload 100 MW power for 99\% of the hours across 20 years for different combinations of capital costs for wind power, solar PV, and lithium-ion battery storage technology, in each of the three states of \textbf{(a)} Karnataka, \textbf{(b)} Gujarat, and \textbf{(c)} Tamil Nadu. The policy range in which each cost estimate falls is depicted by different colours, explained by the legend on the right. All hybrid systems shown in the figure are solar dominated in the wind-solar mix.}
    \label{fig7}
    \vspace{0.5cm}
\end{figure}

Figure \ref{fig7} shows the levelized costs of hybrid systems for different capital costs of wind, solar PV and battery storage, in the three states of Karnataka, Gujarat, and Tamil Nadu. All costs are assuming a lower cost of capital of 2.5\% and shown for providing a baseload generation profile (note that as shown in Figure \ref{fig2}), costs for both flexible and baseload generation are not too dissimilar). Note that while Figure \ref{fig7} only shows the cost of battery storage capacity, total storage system costs including balance of system costs, power costs, and controls are normally twice that of storage capacity costs alone \cite{cole2020cost, mongird20202020}. So a \$75/kWh capacity cost translates to a storage system cost of \$150/kWh in our model. Levelized costs of hybrid systems today, previously shown in Figure \ref{fig2}, are also shown here in the upper left quadrant in each row (state) in Figure \ref{fig7} and correspond to installation costs of \$700/kW and \$1100/kW for solar PV and wind respectively and battery storage capacity costs of \$200/kWh (or total storage system costs of \$400/kWh) \cite{taylor2020irena, cole2020cost, mongird20202020}. At these levels, levelized costs for 99\% availability systems are in the range of \rupee 6-8/kWh as discussed previously in Figure \ref{fig5}. This would require a \$60/ton carbon tax to displace potential new coal generation. 

Figure \ref{fig7} shows that to displace existing coal with 99\% availability hybrid systems across all the three states, without any policy intervention, would require cost targets of at least \$250/kW for solar PV along with battery storage capacity costs of \$100/kWh. This corresponds to a 50\% cost reduction for energy storage and more than a 60\% cost reduction for solar PV. If the objective is to displace coal power by 2040 in line with the goals of the Paris Agreement \cite{cui2019quantifying,iea}, then this cost reduction must be achieved in the next two decades, consistent with an annual year on year cost decline of roughly 6\%. An annual decline of 6\% will also mean that by 2030, hybrid systems will be cheaper than new coal generation in both Karnataka and Tamil Nadu without any economic incentives. 

Note that the results shown in Figure \ref{fig7} correspond to a 2.5\% cost of capital. However, if we consider capital costs of 10\%, even more ambitious cost declines in the next two decades would be necessary for hybrid systems that are cost competitive with coal generation in India. 

Figure \ref{fig7} also shows that the relative resources of solar and wind influence the nature of cost declines for each state. For example, in Gujarat which has the highest share of wind generation in an optimal mix hybrid system, cost reductions are achieved through a fall in costs of any of the three technologies, illustrated by the pattern shown in Figure \ref{fig7}(b). However, for Karnataka and Tamil Nadu, cost declines are mostly influenced by the capital cost of solar PV and storage as in these states wind is a small share of the optimal hybrid system, even in cases where wind capital costs are lower than solar PV (Figure \ref{fig7}(a,c)).

\section*{Discussion}
High capital costs for battery storage and highly seasonal wind generation make it expensive to currently replace coal generation in India with hybrid power plants, even in states with relatively higher coal generation costs. However, falling costs for these technologies together with project financing at lower interest rates can help bridge the gap. Access to capital at low costs, for example through the Green Climate Fund (GCF) or international development assistance, will be critical to expanding the penetration of clean energy in developing countries such as India \cite{chawla2020analysing, hirth2016role, ondraczek2015wacc, schmidt2014low}. Even then, battery storage capacity costs at least 50\% cheaper than current costs and wind and solar PV capital costs that are at least 60\% lower than today's levels will be required to phase out existing coal generation in India without economic incentives for clean energy or carbon taxes. A year on year decline of 6\% in the levelized cost of hybrid systems over the next decade can help avoid the construction of new coal power plants beginning 2030 and achieve phaseout of existing plants by 2040, consistent with the goals of the Paris Agreement. Policy solutions such as allowing for excess wind and solar generation from hybrid plants to be contracted separately as well as technological improvements in low cost grid flexibility mechanisms can help further lower costs and accelerate the timeline of coal phaseout. 

We find that solar PV is more suited to be paired with short duration storage in a hybrid system across the states considered, although blending in some amount of wind capacity can reduce the levelized cost compared to a system with only solar PV and storage. India should adopt a differentiated strategy, focusing first on those states where the variable cost of coal is highest, and gradually use renewable energy with storage to displace coal in other parts of the country as the costs of hybrid systems fall.

\clearpage
\begin{methods}

\subsection{Renewable Energy Data:}
We undertook a three step process for gathering the required multi-year hourly wind and solar generation data for our site locations. 

First, we obtained 20 years of hourly wind and solar output based on MERRA-2 satellite reanalysis data for the period 2000-2019 (both years inclusive). MERRA-2 has a horizontal resolution of 0.5$\degree$ by latitude [−90–90$\degree$] and 0.625$\degree$ by longitude [−180–179.375\degree] with 361 x 576 grid cells worldwide\cite{gelaro2017modern}. We obtained 20 year hourly wind and solar production data based on MERRA-2 for the same exact proprietary data site locations (see first step above) in Karnataka, Gujarat and Tamil Nadu, using the publicly available Renewables.ninja data source by Staffell and Pfenninger which converts weather data from the global reanalysis models and satellite observations to hourly wind and solar output. Solar irradiance data is converted into power output based on the Global Solar Energy Estimator model\cite{pfenninger2016long}. Wind speeds are converted into wind turbine power output using the Virtual Wind Farm model\cite{staffell2016using}.

We use the same wind turbine (hub height and power curves) and solar panel (azimuth angle, tilt angle and tracking selection) configurations as used by the developers at the sites in these states to maintain consistency with our proprietary data.

In the second step, to account for the fact that wind and solar output from reanalysis data such as MERRA-2 can have issues with capacity factors not matching real world observed capacity factors, we scale the median of the 20 years of reanalysis data capacity factors to the observed capacity factors from the proprietary data (Table 1) we have for these states (we scale the MERRA-2 data by the ratio between our proprietary data CF and the median CF of the 20 individual years of MERRA-2 data). This approach, where we draw on the reanalysis data for incorporating the temporal and spatial characteristics of the resource over multiple years, but scale the data to real world observed data for capacity factors (which in our case were obtained at the plant level from industry developers) has been recently outlined in Tong et al 2021\cite{tong2021geophysical}. The section below describes our proprietary data source for observed CFs. The unscaled CFs for the raw MERRA-2 based dataset at the proprietary site locations our shown in Tables S9, S10 and S11 in the SI Note 3.

\subsection{Proprietary Data:}
Proprietary wind and solar power production data normalized to plant capacity were obtained from project operators under Non Disclosure Agreement (NDA) for six power plant sites (one each for wind and solar PV in each state) in the three states of Karnataka, Tamil Nadu, and Gujarat. In each state, the location of the wind and solar PV site was less than 100km from each other to ensure that it could represent a potential co-located hybrid system. Data for Karnataka were obtained at 15 minute intervals for three full years beginning January 1, 2018 up to December 31, 2020. We removed the leap day (29th February) for 2020 from our data. We first averaged across the four 15 minute timestamps to resolve the data at hourly level. Then, we averaged across the three years for each hour of the year (8,760h) to obtain a typical year of hourly normalized production for the wind and solar plant. Some hours of production data were missing in each year but available in the other year. Missing data are due to the SCADA based data collection systems being offline or losing wireless connection and therefore unable to transmit data. For solar PV data in Karnataka, 80\% of the hours of the year (7008/8760) had data in all three years and less than 1\% of hours had data only in one out of the three years. There were no hours with data unavailable across all three years. Therefore, after averaging across the available data for each hour of the year, we obtained a typical year of solar production for use in the optimization model. The same strategy was followed for wind power. 83\% of hours in the year had data in all three years and less than 0.2\% of hours had data only in one out of the three years. Again, there were no hours with data unavailable across all three years. 

The same process was undertaken for the sites in Tamil Nadu. In the case of solar PV, 70\% of hours had data across all three years with 5\% of the hours only having data in one out of the three years. There were no hours with data unavailable across all three years. In the case of wind, 63\% of hours had data only in two out of the three years and 4\% of the hours had data only in one out of the three years. Again, there were no hours with data unavailable across all three years. For Gujarat, normalized wind and solar PV plant level production data with the same level of granularity (15 minute intervals) was available from January 1, 2019 up to April 15, 2021. The rest of the procedure was similar to the other states. For solar power, 15\% of hours had data only in one out of the roughly two and a half years period. There were zero hours with data unavailable across the entire period. For wind, 4\% of hours had data only in one out of the roughly two and a half years. Again, there were zero hours with data unavailable across the entire time period. 

We then calculated the annual capacity factors from the 8760h of hourly wind and solar data for each of the three states. These capacity factors were previously shown in Table 1. We then used these annual capacity factors to scale the MERRA-2 based hourly wind and solar data for the same site locations as described in the section above.

In the absence of any publicly available multi-year wind resource data for India, previous studies\cite{deshmukh2021least, palchak2017greening, gulagi2017electricity} are based on a single year of national wind resource data published by the National Renewable Energy Laboratory (NREL)\cite{nrel}. To our knowledge, this is the first study on India to use multiple years of both wind and solar power production data. Our modelled hybrid plants have an expected lifetime of 20 years so we run our optimization model over the full 20-year period using the entire scaled MERRA-2 based wind and solar dataset.

\subsection{Robustness Checks:}
We performed multiple checks using different available data sources for wind and solar generation to ensure robustness of our results. For more on this see Supplementary Note 3. Overall, we find that our results, i.e. suitability of solar for pairing with lithium-ion battery storage and the high cost of hybrid systems today compared to existing coal plants in India, are both robust to alternate data inputs. 

We also find, consistent with existing literature on the subject\cite{sepulveda2021design, dowling2020role}, that just a small portion of hours in the twenty year dataset are significantly influencing the design of the system and as a result, driving up costs. These results affirm once again the need for multi-year resource datasets for electricity planning.

A detailed exploration of wind and solar resources in the three states is provided in SI Note 4. We find high seasonality for wind generation with significant generation during the monsoon period (June-September) and limited output for the remaining parts of the year. Wind also has notable inter-annual variability in output. Solar output is found to be more consistent both across years and seasons.


\subsection{Coal plant operational profile:}
To simulate the operational profile of coal power plants that hybrid systems will have to meet we considered two separate cases. First, we considered an inflexible baseload profile, operating every hour of the the year at 100 MW.

Second, to simulate flexible generation profiles, we incorporated data on the relative shares of installed wind and solar PV capacity in each state as of February 2021 \cite{mnre2021} as these might influence the shape of the generation profile coal power would be required to produce to balance intermittent renewable energy production in the state. We did this as follows. Let the share of solar PV capacity in state i, defined as the amount of solar PV capacity in the state divided by the sum of wind and solar PV capacity in the state , be S$_{i}$. Similarly, let the wind share be W$_{i}$. We then obtained 20 years of hourly MERRA-2 satellite reanalysis based wind and solar data for the period 2000-2019 (both years inclusive) for four individual locations (North, South, East, West) in each state. Averaging across these four locations provided a potential representation of the average hourly wind and solar resource at the state level for 20 years (175200h). We use the same wind turbine (hub height and power curves) and solar panel (azimuth angle, tilt angle and tracking selection) configurations as previously used for the proprietary site locations. Note that wind and solar resource data for these multiple locations and the state averaged profile were also used in the section on modelling cost impacts of geographically smoothing hybrid systems. CFs for wind and solar in each state (i.e. the state averaged CF) are shown in Table 3 as well as Tables S4 and S5 in the SI.

Let s$_{h,i}$ be the normalized hourly production from solar energy in state i and w$_{h,i}$ be the normalized hourly production from wind energy in state i.

Then we simulate the hourly flexible coal profile for state i (c$_{h,i}$) as:

\begin{equation}
    c_{h,i} = 100 - (100*S_{i} * s_{h,i}) - (100*W_{i} * w_{h,i})
\end{equation}

This gives us a target generation profile for the flexible operation of a coal power plant for 175200h or 20 years. We then use this as the target generation profile for the hybrid power plant. Note that a 24h average of this profile was shown for illustration purposes in Figure \ref{fig1}(b). In our simulation we use the full 175200h hourly data.

Naturally this method of estimating the possible flexible generation profile of a coal plant is a simplification of the actual flexible operation of coal plants in any state. This does not incorporate seasonal dynamics where coal may be offline completely due to high wind production or high hydro power output. Nevertheless using the relative shares of existing intermittent wind and solar capacity for each state allows us to gain some insight into the type of operational profile hybrid systems will be required to fulfil. For example, as Karnataka currently has the highest relative share of solar PV capacity across the three states, its ramp down and up before and after the hours solar PV produces power is more exaggerated than Gujarat or Tamil Nadu (Figure \ref{fig1}(b)). 

Using these simulated baseload and flexible profiles \cite{ziegler2019storage} allows us to then arrive at a bounding estimate of the cost of hybrid systems that may be required to replace coal power using the optimization model described in the next section.

As a further robustness check, we used observed net load data (state wide demand minus state wide intermittent renewable energy) from the year 2018 for Karnataka as a generation profile target in our optimization model. For more details on this see the Limitation Section below and SI Note 5. We found that our estimates of levelized cost for flexible hybrid systems were not strongly driven by the particular shape of our stylized flexible profile.

\subsection{Model:}
We build on the framework described in Ziegler et al \cite{ziegler2019storage}. In this optimization problem, the design objectives consider the capacity of the solar plant, wind plant and battery that must be built; the charging and discharging decision for the battery system at each time step; and the rate of charge or discharge. The objective is to minimize the LCOE of the hybrid system, which is simply the present value of all costs divided by the total energy sold by the system over its lifetime.

The optimization objective is based on the full 20 years of operation (175200 h). The costs are composed of capital costs including installation costs of the battery system, wind plant and solar PV plant. We ignore operating costs such as maintenance given that capital costs far outweigh operating costs \cite{ziegler2019storage, sepulveda2021design}.

This optimization problem is constrained foremost by the target output profile of the coal plant that the hybrid system must match. Further, the operation of the battery is constrained to the maximum rate of charge or discharge, the energy level from one time step to the next, the maximum and minimum energy level. These further bound the problem. 

Our optimization problem is a Linear Programming (LP) problem which is therefore convex and quickly yields globally optimal solutions. The model was run using the General Algebraic Modelling Software (GAMS). The CPLEX solver was used to find globally optimal solutions for the LP problem. We also developed a Mixed Integer Non Linear Programming (MINLP) variation of the model which accounts for the integer dispatch of either charging or discharging of the battery to constrain that the battery cannot both charge and discharge in the same time step. The BARON solver was used to find globally optimal solutions for the MINLP formulation \cite{kilincc2018exploiting,tawarmalani2005polyhedral}. This further specification on the battery yields the exact same solutions as the LP problem as the \textit{net behavior} of the battery in each hour ends up the same across both formulations. However, the MINLP formulation takes significantly longer to solve. We therefore use the LP formulation for running the full 20 year simulation and specify that version below. We also note that given our time step resolution is hourly, it is entirely feasible that storage systems may have both charge and discharge activity within the same one hour period.

The model is described below.

\subsection{Optimization model:}
Let $S_c$, $W_c$, and $B_C$ be the installed capacity size of the solar plant (MW), wind plant (MW) and storage system (MWh) respectively. Let $B_P$ be the rated power capacity of the storage system where $B_P$ = 0.25 * $B_C$ for a four hour storage system. Also let $C_S$, $C_W$, be the total cost of installation per MW for wind and solar, $C_B$ the energy capacity cost of storage per MWh, and $C_P$ the per MW power capacity cost of storage which includes balance of system costs for power equipment, controls, and system integration. Then we have:
\vspace{-0.3cm}
\begin{align}
    C_{\text{system}} = S_{c}*C_{S} + W_{c}*C_{W} + B_{c}*C_{B} + B_{P} * C_{P}
\end{align}

We also apply a capital recovery factor (CRF) = $\frac{r(1+r)^n}{(1+r)^n - 1}$ to the cost of the system, assuming equal yearly payments. N is the plant lifetime and r is the interest rate. We assume a 20 year plant lifetime and three different values for the weighted average cost of capital: 2.5\%, 5\%, and 10\%. Note that the cost of capital is assumed to be the same across all technologies (wind, solar, batteries) and therefore only applies a monotonic transformation on the final levelized cost of the system but does not affect the capacity of different technologies in the optimal mix. Values for all model parameters are summarized in Table \ref{table5}. For sensitivity analysis on certain parameters such as battery storage lifetime, storage roundtrip efficiency, and storage duration (4 hour vs 6, 8 or 10 hour storage) see SI Note 1 (Tables S6-S8). Installation costs for wind and solar PV were obtained from the International Renewable Energy Agency \cite{taylor2020irena} (note that we rounded up cost estimates to the nearest hundred). Costs for battery storage systems are derived from Cole et al and Mongird et al \cite{cole2020cost, mongird20202020}.

The optimization problem minimizes total system cost over the total energy sold by the plant over the lifetime, subject to operating constraints. First we introduce some notation and then describe the optimization setup below. Let $P_{target,t}$ be the hourly target generation profile at time t which is exogenous and was shown earlier in Figure \ref{fig1}. Let $\Delta$t be the one hour time increment. The total energy sold by the hybrid system across the twenty year operational period is then simply the sum of $P_{target,t}$ and is a constant. For example, for the baseload generation case and 100\% availability, the total energy sold is simply 175200h * 100 MW. Let $P_{S,t}$, $P_{W,t}$ be the normalized solar and wind output respectively at time t which is again an exogenous input into the model, from our twenty year hourly wind and solar resource data. Let $P_{B,Ch, t}$ be the battery storage charge behavior at time t and $P_{B,DCh, t}$, the battery storage discharge behavior at time t. Let $\eta$ be the roundtrip efficiency of the storage system, $\eta_{ch}$ be the charge efficiency, and $\eta_{dch}$ the discharge efficiency such that $\sqrt{\eta} = \eta_{ch} = \eta_{dch}$ \cite{hittinger2010compensating}. Let the difference between the the output of the hybrid system and the target output profile $P_{target,t}$ be constrained to always be greater than a pre-defined small tolerance value of $\epsilon$. Finally, let $B_{e,t}$ be the energy level of the battery at time t which is also non-negative. We initialize the model with initial battery storage SOC at 50\%. This is then also set as the final battery storage SOC for the system for the no free energy criterion, i.e. SOC at the first and final time step is conserved. Then we have:

\vspace{-0.2cm}
\begin{mini!}|l|[2]
  {S_c, W_c, B_c, P_{B,Ch,t}, P_{B,DCh,t}, B_{e,t}} {\frac{CRF * C_{\text{system}}}{\sum_1^{T} P_{target,t} \Delta t}}{}{}
  \addConstraint{B_{e,t}}{\leq B_c}
  \addConstraint{B_{e,t}}{=B_{e,t-1}-\frac{1}{\eta_{dch}}*P_{B,DCh,t-1} + \eta_{ch}*P_{B,Ch,t-1}} 
  \addConstraint{P_{B,Ch,t}}{\leq \frac{1}{4} * B_c}
  \addConstraint{P_{B,DCh,t}}{\leq \frac{1}{4} * B_c}
  \addConstraint{P_{B,Ch,t}}{\leq P_{S,t}*S_{c}* + P_{W,t}*W_{c}}
  \addConstraint{\epsilon}{\leq P_{S,t}*S_{c}* + P_{W,t}*W_{c} + P_{B,DCh,t} - P_{B,Ch,t} - P_{target,t}}
  \addConstraint{0.5*B_c}{= B_{e,1}}
  \addConstraint{B_{e,T}}{= B_{e,1}}
  \addConstraint{0}{\leq P_{B,Ch,t}}
  \addConstraint{0}{\leq P_{B,DCh,t}}
  \addConstraint{0}{\leq B_{e,t}}
  \addConstraint{0}{\leq S_{c}}
  \addConstraint{0}{\leq W_{c}}
\end{mini!}

\begin{table}[ht]
    \caption{Model Parameters} 
    \vspace{0.3cm}
    \centering
    \begin{tabular}{l l c c} 
    \textbf{Symbol/Name} & \textbf{Description} & \textbf{Value} & \textbf{Units} \\
     \hline
     $C_S$ & Capital Cost of Solar & 700 & \$/kW \\
    \hline
     $C_W$ & Capital Cost of Wind & 1100 & \$/kW \\
     \hline
     $C_B$ & Capital Cost of Battery Storage Energy Capacity & 200 & \$/kWh \\
     \hline
     $C_P$ & Capital Cost of Battery Storage Power Capacity & 800 & \$/kW \\
    \hline
    Currency Conversion & Value of 1 USD in Rs & 70 &  \\
    \hline
    n & Plant lifetime & 20 & Years\\
    \hline
    $\eta$ & Round Trip Battery Efficiency & 0.75 & -\\
 \hline
\end{tabular}
\label{table5}
\end{table}

\clearpage

\subsection{Coal plant database:}
We used the power plant database assembled by Vasudha Foundation\cite{vasudha} which compiles data from both state and central government sources including state and central regulatory commissions and the Central Electricity Authority (CEA). We only considered operational coal plants and not retired coal plants or coal plants in the pipeline for construction. Total operational coal capacity across the three states was found to be 38,733 MW. Of this capacity, 41\% was privately owned, 38\% was state government owned, and the remaining 21\% was owned by the central government.

\subsection{Limitations:}
Finally, we note a couple of limitations of this work. First, the target coal generation profiles, for both the baseload and flexible cases presented in Figure \ref{fig1}(a,c), are hypothetical. Actual annual generation at coal power plants will differ from this stylized profile. Coal power plants also do not run for all hours of the year and therefore hybrid systems will also be allowed to have downtime. In the absence of plant level data for coal generation we believe our methods to be a useful approximation and that the two profiles can provide a bounding estimate on costs. As an additional robustness check, for Karnataka we collected confidential data from the Karnataka Power Transmission Corporation Limited (KPTCL) of the hourly net load for the state in 2018, which is defined as the hourly electricity demand in the state minus the hourly generation from intermittent renewable energy located in the state. The data were then normalized to a maximum of 100 MW and used as a target coal generation profile in the optimization model. Results are discussed and shown in SI Supplementary Note 5. The optimized system and levelized costs for a hybrid system replacing this profile of generation do not differ significantly from a system replacing our stylized flexible coal profile for Karnataka offering reassurance that our stylized profiles are not driving our results. We do also note that for some individual coal power plants, output is reduced to almost zero during the monsoon months when wind and hydro power generation is at maximum levels. Incorporating these dynamics would likely further skew the results in favor of combining solar PV with short duration storage as no coal generation would need to be replaced when wind is generating at full power.

Second, we only consider short duration storage in the form of lithium-ion batteries and do not consider long duration ($>$10h) storage solutions. Long duration storage technologies such as flow batteries, thermal storage solutions, or hydrogen based storage are yet to mature and be deployed at scale anywhere in the world. Short-duration storage accounts for approximately 93\% of total global storage power capacity\cite{doestorage}. We believe that focusing on mature technologies that are available today, have already been deployed in India, and have seen consistent cost declines, would be more valuable to researchers, policymakers, and industry actors. Nevertheless, we note that low cost long duration storage technologies could play a valuable role in decarbonizing electricity \cite{sepulveda2021design}. Availability of such technologies in India could hasten the pace of coal phaseout and merit further research.

\end{methods}


\clearpage

\section*{References}
\bibliography{citations.bib}

\clearpage

\begin{addendum}
 \item [Contributions] A.M., S.S., P.V., R.T., A.A., and I.A. developed the concept and designed the paper. A.M. and P.V. designed the code and A.M. undertook model simulations. A.M., S.S., P.V., R.T., A.A., and I.A. analysed the data. A.M. and P.V. designed the Figures and wrote the manuscript on which S.S., R.T., A.A., and I.A. commented.
 \item [Code availability] Code used for the analysis is available from the corresponding author on request.
 \item[Competing Interests] The authors declare that they have no competing financial interests.
 \item[Correspondence] Correspondence and requests for materials should be addressed to A.M. (aniruddh@cmu.edu)
\end{addendum}

\end{document}


\maketitle

\begin{small}
\begin{affiliations}
 \item Department of Engineering and Public Policy, Carnegie Mellon University, Pittsburgh, 15213, USA
 \item School for Environment and Sustainability, University of Michigan, Ann Arbor, 48109, USA
 \item Centre for Social and Economic Progress, New Delhi, 110021, India 
 \item Reconnect Energy, Bangalore, 560008, India
 \item Department of Energy Resources Engineering, Stanford University, Stanford, 94305, USA
\end{affiliations}
\end{small}
\vspace{-0.5cm}

\noindent \textbf{This PDF includes:}\\
Supplementary Notes 1-5.\\
Supplementary Figures 1-7.\\
Supplementary Tables 1-15.

\newpage

\section{Supplementary Note 1: Additional results}

\subsection{Mixing wind and solar PV:}
Figure S1 below shows the change in the LCOE for different mixes of wind and solar PV capacity in the hybrid system for supplying flexible generation throughout the year in each state. The equivalent result for baseload generation was shown in the main manuscript text in Figure 3.

\begin{figure}[!h]
    \vspace{1cm}
    \centering
    \vspace{-0.5cm}
    \includegraphics[width=16cm, height = 5.5cm]{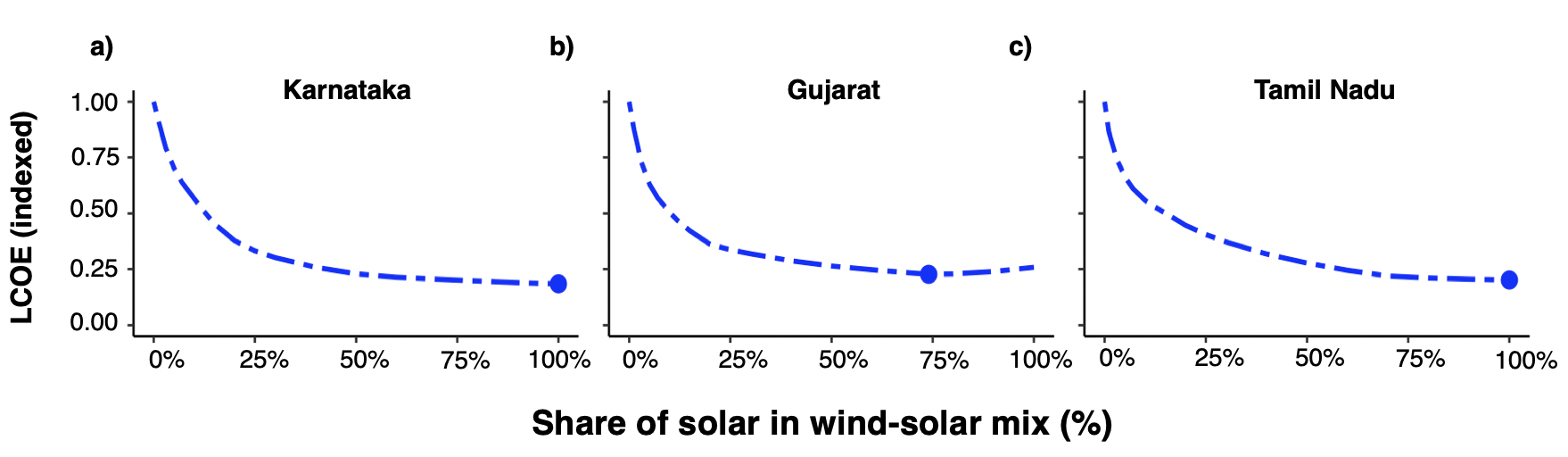}
    \caption{LCOE for different shares of solar PV in hybrid systems with both wind and solar PV providing flexible generation in the three states. LCOE values on the Y axis are indexed to the LCOE for a hybrid system with only wind power and battery storage. This is the flexible generation equivalent to the results shown for baseload generation in Figure 3 of the main manuscript text. \textbf{a)} Results for Karnataka. The optimal mix shown with the blue dot has 100\% solar PV and 0\% wind capacity. \textbf{b)} Results for Gujarat. The optimal mix shown with the blue dot has 74\% solar PV and 26\% wind capacity.  and \textbf{c)} Results for Tamil Nadu. The optimal mix shown with the blue dot has 100\% solar PV and 0\% wind capacity. }\label{fig1}
    \vspace{1cm}
\end{figure}

\subsection{Battery storage operation:}
Figures S2, S3 and S4 show the hourly operation of battery storage throughout the twenty years for Karnataka, Gujarat and Tamil Nadu, across hybrid systems with an optimal mix of wind and solar with storage, solar PV with storage, and wind with storage.

\begin{figure}[!h]
    \centering
    \includegraphics[width=15cm, height = 8cm]{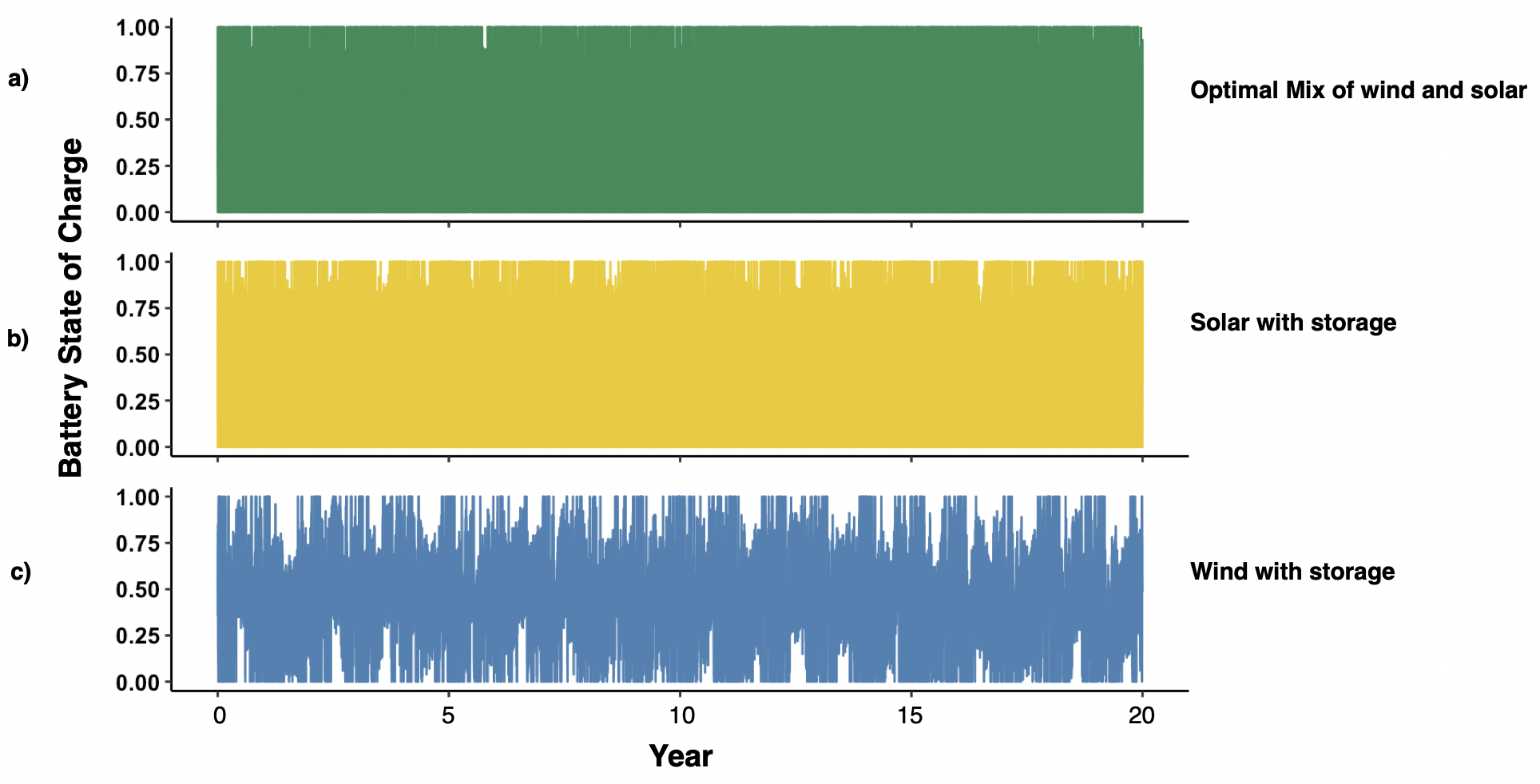}
    \caption{Hourly State of Charge of the battery storage system across 20 years for a hybrid system providing baseload 100 MW generation in \textbf{Karnataka}. \textbf{a)} Hybrid system with an optimal mix of wind, solar PV and battery storage. The individual capacities are 1685 MW solar, 14 MW wind and 2727 MWh of battery storage. \textbf{b)} Hybrid system consisting of solar PV and battery storage. The individual capacities are 1852 MW solar, 0 MW wind and 2595 MWh of battery storage. \textbf{c)} Hybrid system with wind and battery storage. The individual capacities are 0 MW solar, 5500 MW wind and 16626 MWh of battery storage.}\label{fig2}
\end{figure} 

\begin{figure}[!h]
    \centering
    \vspace{-0.5cm}
    \includegraphics[width=15cm, height = 8cm]{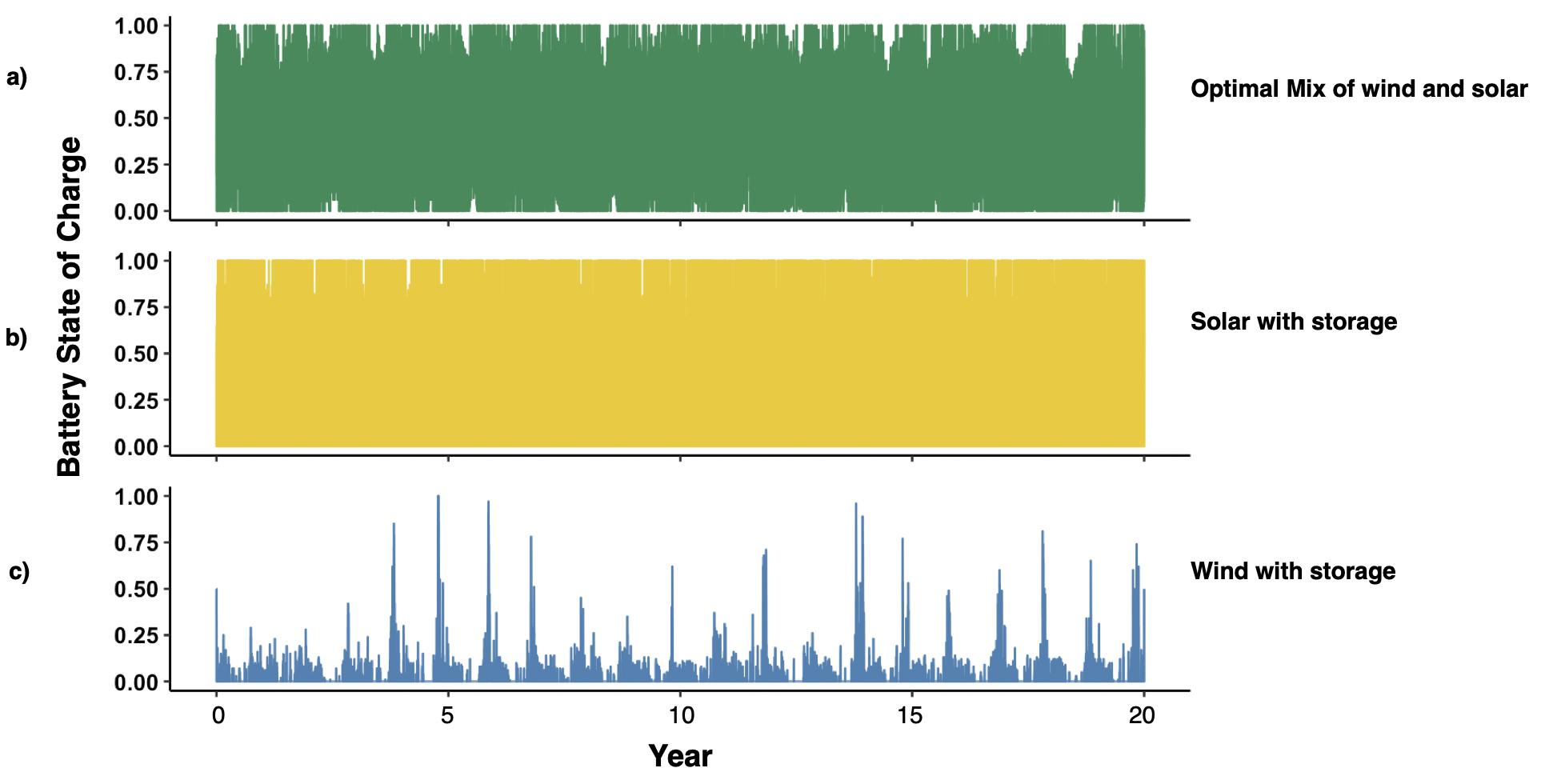}
    \caption{Hourly State of Charge of the battery storage system across 20 years for a hybrid system providing baseload 100 MW generation in \textbf{Gujarat}. \textbf{a)} Hybrid system with an optimal mix of wind, solar PV and battery storage. The individual capacities are 1136 MW solar, 371 MW wind and 2038 MWh of battery storage. \textbf{b)} Hybrid system consisting of solar PV and battery storage. The individual capacities are 2118 MW solar, 0 MW wind and 2668 MWh of battery storage. \textbf{c)} Hybrid system with wind and battery storage. The individual capacities are 0 MW solar, 5544 MW wind and 8503 MWh of battery storage.}\label{fig3}
\end{figure} 

\begin{figure}[!h]
    \centering
    \vspace{-0.5cm}
    \includegraphics[width=15cm, height = 8cm]{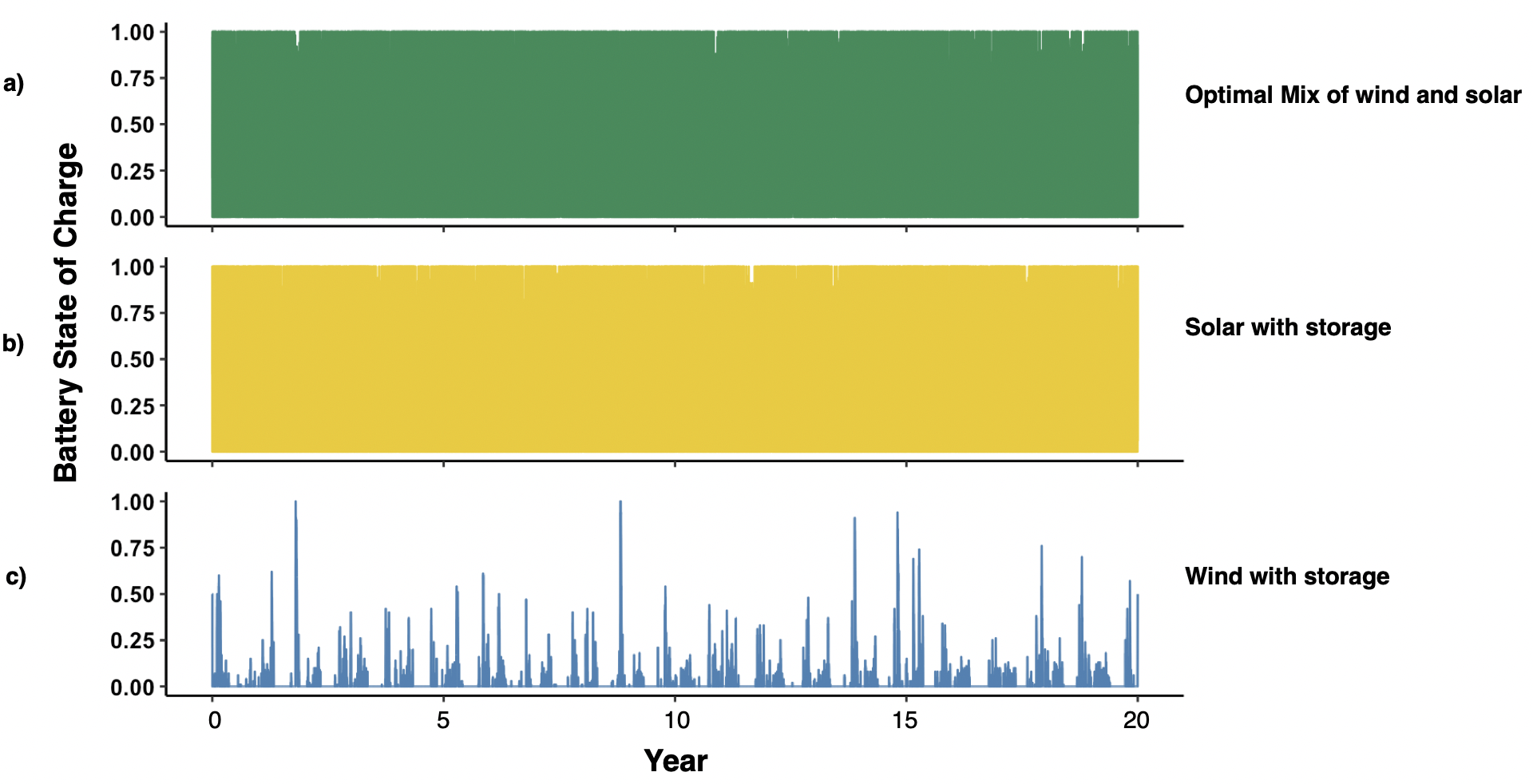}
    \caption{Hourly State of Charge of the battery storage system across 20 years for a hybrid system providing baseload 100 MW generation in \textbf{Tamil Nadu}. \textbf{a)} Hybrid system with an optimal mix of wind, solar PV and battery storage. The individual capacities are 1550 MW solar, 109 MW wind and 2606 MWh of battery storage. \textbf{b)} Hybrid system consisting of solar PV and battery storage. The individual capacities are 1854 MW solar, 0 MW wind and 2617 MWh of battery storage. \textbf{c)} Hybrid system with wind and battery storage. The individual capacities are 0 MW solar, 4911 MW wind and 14387 MWh of battery storage.}
    \label{fig4}
\end{figure}

\clearpage

\subsection{Value of curtailment:}
We consider the case where some economic value is assigned to curtailed energy from the oversized hybrid system, as discussed in the main manuscript text. Figure S7 below shows the change in the LCOE as different levels of economic value as a share of the baseload generation is assigned to excess generation produced by the hybrid system. LCOE values on the Y axis are indexed to the case where no value is assigned to excess generation and corresponds to the LCOE results shown in Figure 1(b) and Figure 1(d) in the main manuscript text. As can be seen, contracting of excess generation can significantly reduce the costs of hybrid systems and help bridge the gap to current costs of coal generation.

\begin{figure}[!h]
    \vspace{1cm}
    \centering
    \vspace{-0.5cm}
    \includegraphics[width=\textwidth, height = 5cm]{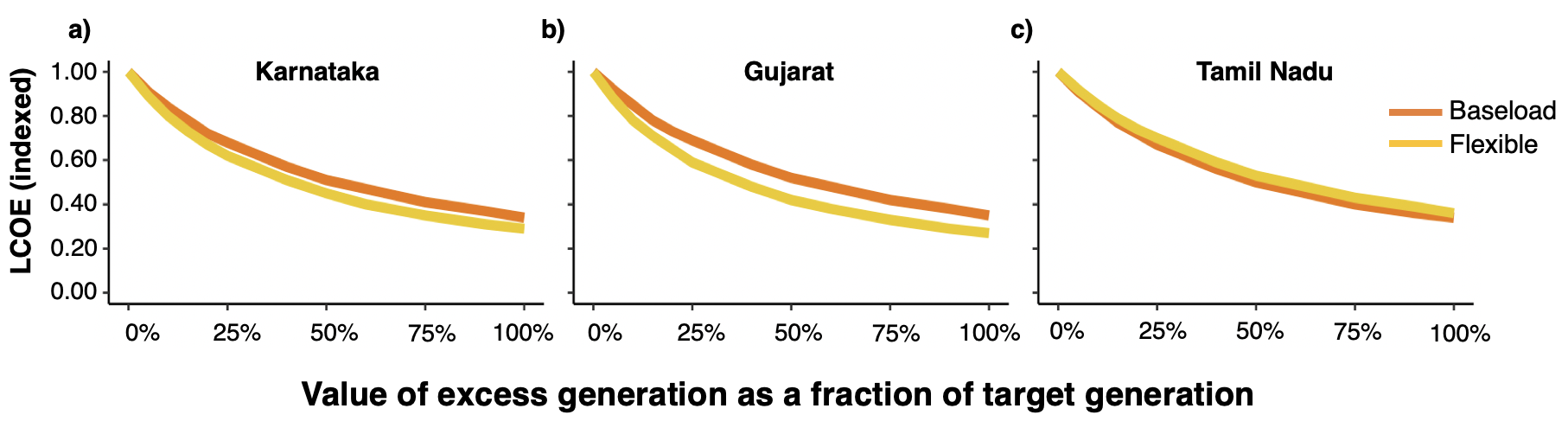}
    \caption{\small{Change in LCOE for different valuations of the curtailed energy from excess wind and solar power generation in the optimal hybrid power plant for each state. LCOE values on the Y axis are indexed to the case where no value is assigned to excess generation which is the default case also presented in the LCOE results shown in the main manuscript. \textbf{a)} Results for Karnataka, curtailment is 190\% in the case of baseload generation and 245\% for flexible generation. \textbf{b)} Results for Gujarat, curtailment is 183\% in the case of baseload generation and 275\% for flexible generation. \textbf{c)} Results for Tamil Nadu, curtailment is 197\% in the case of baseload generation and 174\% for flexible generation.}}
    \label{fig5}
\end{figure}
\clearpage

\subsection{Reduced Availability \\}
Tables S1-S3 shows the capacity and costs of hybrid systems providing baseload 100 MW power for 20 years across different availability requirements for the three states. Note that Figure 5 in the main manuscript text also showed the levelized cost estimates. Tables S1-S3 show that across all three states, as the availability required is lowered, levelized costs fall and the share of solar in the wind-solar mix also drops. Reductions in costs are achieved by reduction in the size of storage capacity required, and wind begins to substitute for storage capacity in the optimal mix.

\begin{table}[!ht]
\begin{center}
\caption{\textbf{\small Capacity and levelized costs of hybrid systems providing 100 MW baseload power for 20 years across different availability requirements for Karnataka. All LCOE values assume a 2.5\% cost of capital and a fixed exchange rate of 1USD = \rupee70.}}
\begin{tabular}{r|r|r|r|r|r|r} \textbf{Availability} & \textbf{\begin{tabular}[c]{@{}l@{}}LCOE \\ (\$/MWh)\end{tabular}} & \textbf{\begin{tabular}[c]{@{}l@{}}LCOE \\ (Rs/kWh)\end{tabular}} & \textbf{\begin{tabular}[c]{@{}l@{}}Solar \\ (MW)\end{tabular}} & \textbf{\begin{tabular}[c]{@{}l@{}}Wind \\ (MWh)\end{tabular}} & \textbf{\begin{tabular}[c]{@{}l@{}}Storage \\ (MWh)\end{tabular}} & \textbf{\begin{tabular}[c]{@{}l@{}}Wind \% in \\ wind-solar mix\end{tabular}} \\
\hline
100\%                 & 167.0                                                             & 11.7                                                              & 1685                                                           & 14                                                             & 2727                                                              & 1\%                                                                            \\
\hline
                    99\%                  & 99.5                                                              & 7.0                                                               & 916                                                            & 105                                                            & 1479                                                              & 10\%                                                                           \\
\hline
                    95\%                  & 82.4                                                              & 5.8                                                               & 609                                                            & 138                                                            & 1235                                                              & 18\%                                                                           \\
\hline
                    90\%                  & 73.1                                                              & 5.1                                                               & 447                                                            & 215                                                            & 879                                                               & 32\%                                                                           \\
\hline
                    85\%                  & 65.6                                                              & 4.6                                                               & 346                                                            & 262                                                            & 581                                                               & 43\%                                                                           \\
\hline
                    80\%                  & 58.8                                                              & 4.1                                                               & 274                                                            & 285                                                            & 346                                                               & 51\%  
\end{tabular}
\end{center}
\end{table}

\begin{table}[!hbt]
\begin{center}
\caption{\textbf{\small Capacity and levelized costs of hybrid systems providing 100 MW baseload power for 20 years across different availability requirements for Tamil Nadu. All LCOE values assume a 2.5\% cost of capital and a fixed exchange rate of 1USD = \rupee70.}}
\begin{tabular}{r|r|r|r|r|r|r}
\textbf{Availability} & \textbf{\begin{tabular}[c]{@{}l@{}}LCOE \\ (\$/MWh)\end{tabular}} & \textbf{\begin{tabular}[c]{@{}l@{}}LCOE \\ (Rs/kWh)\end{tabular}} & \textbf{\begin{tabular}[c]{@{}l@{}}Solar \\ (MW)\end{tabular}} & \textbf{\begin{tabular}[c]{@{}l@{}}Wind \\ (MWh)\end{tabular}} & \textbf{\begin{tabular}[c]{@{}l@{}}Storage \\ (MWh)\end{tabular}} & \textbf{\begin{tabular}[c]{@{}l@{}}Wind \% in \\ wind-solar mix\end{tabular}} \\
\hline
100\%                 & 164.1                                                             & 11.5                                                              & 1550                                                           & 109                                                            & 2606                                                              & 7\%                                                                            \\
\hline
                     99\%                  & 105.0                                                              & 7.3                                                               & 1029                                                           & 92                                                             & 1504                                                              & 8\%                                                                            \\
\hline
                    95\%                  & 86.7                                                              & 6.1                                                               & 662                                                            & 128                                                            & 1305                                                              & 16\%                                                                           \\
\hline
                    90\%                  & 78.1                                                              & 5.5                                                               & 479                                                            & 205                                                            & 1003                                                              & 30\%                                                                           \\
\hline
                    85\%                  & 71.3                                                              & 5.0                                                               & 363                                                            & 268                                                            & 704                                                               & 42\%                                                                           \\
\hline
                    80\%                  & 65.3                                                              & 4.6                                                               & 282                                                            & 301                                                            & 467                                                               & 52\%                
\end{tabular}
\end{center}
\end{table}

\begin{table}[!hbt]
\begin{center}
\caption{\textbf{\small Capacity and levelized costs of hybrid systems providing 100 MW baseload power for 20 years across different availability requirements for Gujarat. All LCOE values assume a 2.5\% cost of capital and a fixed exchange rate of 1USD = \rupee70.}}
\begin{tabular}{r|r|r|r|r|r|r}
\textbf{Availability} & \textbf{\begin{tabular}[c]{@{}l@{}}LCOE \\ (\$/MWh)\end{tabular}} & \textbf{\begin{tabular}[c]{@{}l@{}}LCOE \\ (Rs/kWh)\end{tabular}} & \textbf{\begin{tabular}[c]{@{}l@{}}Solar \\ (MW)\end{tabular}} & \textbf{\begin{tabular}[c]{@{}l@{}}Wind \\ (MWh)\end{tabular}} & \textbf{\begin{tabular}[c]{@{}l@{}}Storage \\ (MWh)\end{tabular}} & \textbf{\begin{tabular}[c]{@{}l@{}}Wind \% in \\ wind-solar mix\end{tabular}} \\
\hline
100\%                 & 147.5                                                             & 10.3                                                              & 1136                                                           & 371                                                            & 2038                                                              & 25\%                                                                           \\
\hline
                    99\%                  & 88.4                                                              & 6.2                                                               & 615                                                            & 191                                                            & 1397                                                              & 24\%                                                                           \\
\hline
                    95\%                  & 67.4                                                              & 4.7                                                               & 370                                                            & 295                                                            & 727                                                               & 44\%                                                                           \\
\hline
                    90\%                  & 53.6                                                              & 3.8                                                               & 245                                                            & 349                                                            & 258                                                               & 59\%                                                                           \\
\hline
                    85\%                  & 45.8                                                              & 3.2                                                               & 200                                                            & 318                                                            & 106                                                               & 61\%                                                                           \\
\hline
                    80\%                  & 41.0                                                              & 2.9                                                               & 177                                                            & 284                                                            & 29                                                                & 62\% \\
\end{tabular}
\end{center}
\end{table}

\clearpage
\subsection{Geographical smoothing and cost of hybrid systems in alternate locations\\}
Here we consider the effect of smoothing wind and solar resources over multiple locations in a state, in order to potentially reduce the cost of hybrid systems. The main manuscript text (Table 3) presented the results for Gujarat. In Table S4 below, we show the results for the state of Karnataka for replacing 100 MW baseload power for twenty years using hybrid systems located at individual sites in the North, East, South, West, and finally based on the state averaged resource which averages wind and solar output across the four individual locations. Wind and solar resource data is based on the MERRA-2 reanalysis dataset as explained in Methods in the main manuscript text. Note that in this case we use the raw, unscaled data, as we do not have observed capacity factor data for these locations.

\begin{table}[hbt!]
\caption{\textbf{\small Levelized costs and capacity mix of hybrid power systems providing 100 MW baseload power over 20 years in the state of Karnataka at different individual site locations and a state averaged location which averages out resources from the four individual locations. All levelized costs shown are based on a 2.5\% cost of capital.}}
\begin{tabular}{l|r|r|r|r|r|r|r}
\label{table1}

\textbf{Location} & \textbf{Solar CF} & \textbf{Wind CF} & \textbf{\begin{tabular}[c]{@{}l@{}}Solar \\ (MW)\end{tabular}} & \textbf{\begin{tabular}[c]{@{}l@{}}Wind \\ (MW)\end{tabular}} & \textbf{\begin{tabular}[c]{@{}l@{}}Battery \\ (MWh)\end{tabular}} & \textbf{\begin{tabular}[c]{@{}l@{}}Solar \% in \\ least cost mix\end{tabular}} & \textbf{\begin{tabular}[c]{@{}l@{}}LCOE \\ (Rs/kWh)\end{tabular}} \\
\hline
North      & 22\%     & 26\%    & 1455                                                  & 119                                                  & 2018                                                     & 92\%                                                                  & 10.0                                                     \\
\hline
East       & 22\%     & 30\%    & 1444                                                  & 100                                                  & 2269                                                     & 94\%                                                                  & 10.4                                                     \\
\hline
South      & 21\%     & 13\%    & 1777                                                  & 0                                                    & 2687                                                     & 100\%                                                                 & 11.9                                                     \\
\hline
West       & 21\%     & 14\%    & 1733                                                  & 0                                                    & 2714                                                     & 100\%                                                                 & 11.8                                                     \\
\hline
State avg. & 21\%     & 21\%    & 1153                                                  & 25                                                   & 2073                                                     & 98\%                                                                  & 8.5                                                     
\end{tabular}
\end{table}

Next, Table S5 below shows the equivalent results for the state of Tamil Nadu.

\begin{table}[hbt!]
\caption{\textbf{\small Levelized costs and capacity mix of hybrid power systems providing 100 MW baseload power over 20 years in the state of Tamil Nadu at different individual site locations and a state averaged location which averages out resources from the four individual locations. All levelized costs shown are based on a 2.5\% cost of capital.}}
\begin{tabular}{l|r|r|r|r|r|r|r}

\label{table2}
\textbf{Location} & \textbf{Solar CF} & \textbf{Wind CF} & \textbf{\begin{tabular}[c]{@{}l@{}}Solar \\ (MW)\end{tabular}} & \textbf{\begin{tabular}[c]{@{}l@{}}Wind \\ (MW)\end{tabular}} & \textbf{\begin{tabular}[c]{@{}l@{}}Battery \\ (MWh)\end{tabular}} & \textbf{\begin{tabular}[c]{@{}l@{}}Solar \% \\ in least cost mix\end{tabular}} & \textbf{\begin{tabular}[c]{@{}l@{}}LCOE \\ (Rs/kWh)\end{tabular}} \\
\hline
North             & 21\%              & 28\%             & 1256                                                           & 65                                                            & 2809                                                              & 95\%                                                                           & 10.6                                                              \\
\hline
East              & 18\%              & 36\%             & 1226                                                           & 240                                                           & 2220                                                              & 84\%                                                                           & 10.3                                                              \\
\hline
South             & 18\%              & 35\%             & 1586                                                           & 160                                                           & 2090                                                              & 91\%                                                                           & 10.9                                                              \\
\hline
West              & 20\%              & 21\%             & 1595                                                           & 30                                                            & 2782                                                              & 98\%                                                                           & 11.6                                                              \\
\hline
State avg.        & 19\%              & 30\%             & 881                                                            & 450                                                           & 1523                                                              & 66\%                                                                           & 8.8                                                              
\end{tabular}
\end{table}

We see that for both states, hybrid systems that draw from wind and solar resources from multiple locations are cheaper than the cheapest individual location. For both Tamil Nadu and Karnataka, the state average resource based hybrid system is roughly 15\% cheaper than the cheapest individual location. Drawing from wind and solar resources across multiple locations does not necessarily lead to larger shares for wind energy in the wind-solar mix. Across all three states (main manuscript Table 3, and Table S4-S5) hybrid systems continue to be solar dominated at all locations, including the state average system. Only in the case of Tamil Nadu, where wind rises to 34\% of the mix for the state averaged system, is the share of wind appreciably higher than in optimal systems at any of the individual locations.

Figure S6 shows the LCOE assuming a 2.5\% cost of capital across these different modelled locations, shown on a map of India, as well as the state averaged results. These results also show that the range of levelized cost estimates from our optimization model for hybrid systems are robust to the selection of the site locations across these states.

\begin{figure}[!h]
    \centering
    \includegraphics[width=9cm, height = 9cm]{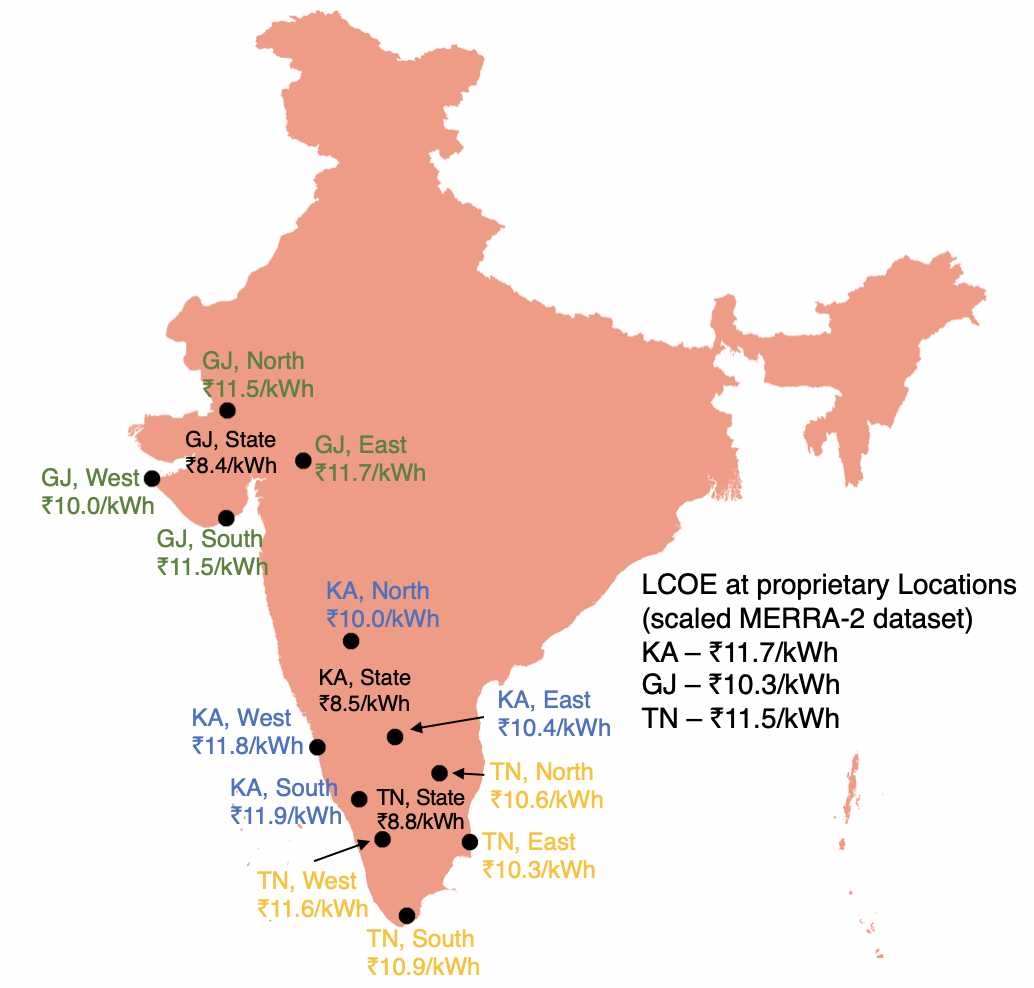}
    \caption{\textbf{\small Map of India showing the LCOE at different site locations in the states of Karnataka, Gujarat, and Tamil Nadu and finally a state averaged hybrid system cost which draws on solar and wind resources from all four locations. Costs are for a least cost hybrid system mixing solar PV, wind power, and lithium-ion batterty storage to provide baseload 100 MW of power throughout the year and are based on a 2.5\% cost of capital. LCOE values at proprietary locations in the three states are listed on the right and are based on the scaled MERRA-2 dataset.}}
    \label{fig6}
\end{figure}
\clearpage

\subsection{Parametric Sensitivity Analysis\\}
Here we present additional results from the sensitivity analysis where we varied certain parameters and assumptions regarding our optimization analysis. We present these results for the state of Gujarat and for the case of generating 100 MW baseload power for 20 years for the sake of brevity, results for other states follow a similar pattern.

\subsubsection*{1. Storage duration: 6, 8, 10 hour systems}
Our main manuscript text results focused on 4 hour duration li-ion battery storage systems. However it is possible that lithium-ion storage could have slower charge/discharge duration, leading to 6, 8, or 10 hour storage. In Table S6 below we show the results for the least cost hybrid system based on longer storage duration and compare them with the baseline results for 4 hour storage that we presented previously. Cost data for 6, 8, and 10 hour storage systems were obtained from Mongird et al \cite{mongird20202020}.

\begin{table}[hbt]
\caption{\textbf{\small Cost and capacity of optimal hybrid systems in the state of Gujarat for providing 100 MW baseload generation across different assumptions around the battery storage duration. Results in the main manuscript are based on 4 hour storage. All LCOE values are based on a 2.5\% cost of capital.}}
\begin{center}
\begin{tabular}{l|r|r|r|r|r}
\textbf{Storage duration} & \textbf{LCOE} & \textbf{\begin{tabular}[c]{@{}l@{}}Solar \\ (MW)\end{tabular}} & \textbf{\begin{tabular}[c]{@{}l@{}}Wind \\ (MW)\end{tabular}} & \textbf{\begin{tabular}[c]{@{}l@{}}Battery \\ (MWh)\end{tabular}} & \textbf{\begin{tabular}[c]{@{}l@{}}Solar (\%) in \\ wind-solar mix\end{tabular}} \\
\hline
4 hour                    & 10.3          & 1136                                                           & 371                                                           & 2038                                                              & 75\%                                                                             \\
\hline
6 hour                    & 10.1          & 1136                                                           & 371                                                           & 2038                                                              & 75\%                                                                             \\
\hline
8 hour                    & 10.0          & 1152                                                           & 368                                                           & 2023                                                              & 76\%                                                                             \\
\hline
10 hour                   & 10.2          & 1258                                                           & 311                                                           & 2124                                                              & 80\%                                                         
\end{tabular}
\end{center} 
\end{table}

As can be seen in Table S6, levelized cost and capacities of hybrid systems are very similar across different assumptions for storage duration indicating that our model results are not sensitive to this parameter. We also see that the share of solar in the wind-solar mix increases slightly as duration of storage increases, indicating that slower discharge systems may be even more conducive to pairing with solar over wind compared to shorter duration storage.

\subsubsection*{2. Storage lifetime: 15 years}
In the baseline results presented in the main manuscript text we assume a storage lifetime of 20 years. Here we reduce the expected lifetime to 15 years with a requirement to replace the storage system with a new capital purchase after 15 years of operation. The remaining plant elements continue to have a 20 year lifetime and note that we do not change the objectives, parameters, or constraints of the optimization but rather, add to the capital cost of the existing optimal system, an additional cost of purchasing a new storage system after 15 years. We assume that total storage system costs drop from \$400/kWh today to \$250/kWh 15 years from now and we consider an 8\% discount rate given higher interest rates in developing countries such as India compared to industrialized nations. Table S7 shows the comparison to the baseline case where we had assume 20 year lifetime. Clearly, costs rise if the storage system needs to be replaced sooner, but this increase is not significant. 

\begin{table}[]
\caption{\textbf{\small Cost and capacity of optimal hybrid systems in the state of Gujarat for providing 100 MW baseload generation across different assumptions around the battery storage lifetime. Note that we do not change the objectives, parameters or constraints in the optimization model but only add an additional cost of replacing the storage system sooner. Results in the main manuscript are based on 20 year lifetime. All LCOE values are based on a 2.5\% cost of capital.}}
\begin{center}
\begin{tabular}{l|r|r|r|r}
\textbf{Lifetime} & \textbf{LCOE} & \textbf{\begin{tabular}[c]{@{}l@{}}Solar \\ (MW)\end{tabular}} & \textbf{\begin{tabular}[c]{@{}l@{}}Wind \\ (MW)\end{tabular}} & \textbf{\begin{tabular}[c]{@{}l@{}}Battery \\ (MWh)\end{tabular}} \\
\hline
20 years & 10.3 & 1136                                                  & 371                                                  & 2038                                                     \\
\hline
15 years & 11.1 & 1136                                                  & 371                                                  & 2038                                                    
\end{tabular}
\end{center}
\end{table}

\subsubsection*{3. Storage roundtrip efficiency}
In our baseline model results presented in the main manuscript text we assume a roundtrip efficiency of 75\% and individual charge and discharge efficiency of 87\%, which is in line with other studies\cite{ziegler2019storage}. Here, we run an additional case with higher roundtrip efficiencies, e.g. 85\% and 90\% by simply changing the parameter in the optimization model.

\begin{table}[hbt]
\begin{center}
\caption{\textbf{\small Cost and capacity of optimal hybrid systems in the state of Gujarat for providing 100 MW baseload generation across different assumptions around the battery storage roundtrip efficiency. Results in the main manuscript are based on 75\% roundtrip efficiency. All LCOE values are based on a 2.5\% cost of capital.}}
\begin{tabular}{l|r|r|r|r|r}
\textbf{Roundtrip  efficiency} & \textbf{LCOE} & \textbf{\begin{tabular}[c]{@{}l@{}}Solar \\ (MW)\end{tabular}} & \textbf{\begin{tabular}[c]{@{}l@{}}Wind \\ (MW)\end{tabular}} & \textbf{\begin{tabular}[c]{@{}l@{}}Battery \\ (MWh)\end{tabular}} & \textbf{\begin{tabular}[c]{@{}l@{}}Solar (\%) in \\ wind-solar mix\end{tabular}} \\
\hline
0.75                           & 10.3          & 1136                                                           & 371                                                           & 2038                                                              & 75\%                                                                             \\
\hline
0.85                           & 9.9           & 1061                                                           & 382                                                           & 1934                                                              & 74\%                                                                             \\
\hline
0.90                           & 9.7           & 1020                                                           & 387                                                           & 1888                                                              & 72\%                                                                            
\end{tabular}
\end{center}
\end{table}

Naturally, as can be seen in Table S8, higher roundtrip efficiencies reduce the levelized cost of hybrid systems. We also see that the share of solar in the wind-solar mix drops marginally as efficiency improves, indicating that higher efficiency storage could increase the role for wind in hybrid power plants. Overall, improved efficiencies of storage systems in the future can accelerate the deployment of hybrid power plants on the electricity grid by reducing costs and helping further integrate renewable energy.

\clearpage
\section{Supplementary Note 2: Value of avoided premature mortality from local air pollution}
Particulate matter pollution from coal power generation in India results in significant health impacts through premature mortality. For example, Cropper et al \cite{cropper2021mortality} estimated that in 2018, ambient PM$_{2.5}$ pollution from coal generation in India caused 78,000 premature mortalities. Cropper et al \cite{cropper2021mortality} accordingly estimate the health damages from coal generation in India to be \rupee 0.73/kWh based on a value of statistical (VSL) estimate for India of \rupee 10.3 million for 2018. Even if these health benefits of avoided coal generation are factored into the levelized costs of hybrid power systems presented in Figure 2 in the main manuscript text, this reduction on its own is insufficient to make wind and solar power paired with battery storage competitive with baseload coal generation. We also note that given the scale of cost reductions required for hybrid power plants to be competitive with more expensive coal generation, this result is robust to alternative, higher, specifications of the VSL in India.

\newpage
\section{Supplementary Note 3: Robustness checks with varying wind and solar resource data}
We performed multiple robustness checks using different available data sources for wind and solar generation to ensure robustness of our results.

First, we ran our optimization model using the raw 20 year MERRA-2 based hourly wind and solar generation data, i.e. these data were not scaled to our observed proprietary data capacity factors. Second, we ran our optimization model using only the one year (8760h) of proprietary data for the three states. A description of the proprietary data is provided in the main manuscript text in Methods. Third, we ran our optimization model using publicly available wind and solar resource data from NREL\cite{nrel, sam} which has been the basis of previous studies on India's electricity mix. In the case of wind, we first convert wind speeds from the 2014 NREL dataset\cite{nrel} for wind speeds in India to normalized power production values using the power curves for the respective wind turbines at the three sites. For solar PV, NREL's System Advisor Model\cite{sam} provides hourly estimated production of a solar PV plant for a typical year for user selected locations in India based based on sixteen years of solar irradiance data. We run the model for our solar PV site locations in the three states to generate the hourly production, using the same tracking and tilt parameters as with our proprietary data and MERRA-2 based dataset. The optimization model is then run for a full year using these wind and solar production data.

Tables S9-S11 below show the resource characteristics and results of the optimization model for these different data inputs for hourly wind and solar generation at our proprietary site location. As can be seen, capacity factors (CFs) for wind from the NREL dataset are significantly higher than those estimated from an average year of our proprietary data or from the MERRA-2 based dataset. For solar PV, the CFs are fairly consistent across the different data sources.

\begin{table}[ht]
\caption{\textbf{\small Resource characteristics and optimal hybrid system that provides 100 MW baseload power every hour of the year in the state of Karnataka across different data sources for wind and solar generation. LCOE values are calculated assuming a 2.5\% cost of capital.}}
\begin{tabular}{|l|l|r|r|r|r|r|r|r|r|}
\hline
\textbf{State} & \textbf{\begin{tabular}[c]{@{}l@{}}Data \\ Source\end{tabular}} & \textbf{\begin{tabular}[c]{@{}l@{}}Wind \\ CF\end{tabular}} & \textbf{\begin{tabular}[c]{@{}l@{}}Solar \\ CF\end{tabular}} & \textbf{Years} & \textbf{\begin{tabular}[c]{@{}l@{}}Solar \\ (MW)\end{tabular}} & \textbf{\begin{tabular}[c]{@{}l@{}}Wind \\ (MW)\end{tabular}} & \textbf{\begin{tabular}[c]{@{}l@{}}Battery \\ (MWh)\end{tabular}} & \textbf{\begin{tabular}[c]{@{}l@{}}Solar (\%) in \\ wind-solar \\ mix\end{tabular}} & \textbf{\begin{tabular}[c]{@{}l@{}}LCOE \\ (Rs/kWh)\end{tabular}} \\
\hline
KA             & \begin{tabular}[c]{@{}l@{}}MERRA-2 \\ (Unscaled)\end{tabular} & 22\%             & 22\%              & 20                                                                      & 1449                                                           & 14                                                            & 2726                                                              & 99\%                                                                             & 10.8                                                              \\
\hline
KA             & \begin{tabular}[c]{@{}l@{}}MERRA-2 \\ (Scaled)\end{tabular}   & 22\%             & 19\%              & 20                                                                      & 1685                                                           & 14                                                            & 2727                                                              & 99\%                                                                             & 11.7                                                              \\
\hline
KA             & Proprietary                                                   & 22\%             & 19\%              & 1                                                                       & 745                                                            & 112                                                           & 1593                                                              & 87\%                                                                             & 6.6                                                               \\
\hline
KA             & NREL                                                          & 25\%             & 21\%              & 1                                                                       & 882                                                            & 182                                                           & 1615                                                              & 83\%                                                                             & 7.5                                                               \\
\hline         
\end{tabular}
\end{table}

We find that the hybrid systems based on the unscaled MERRA-2 dataset are similar in capacity mix and cost to the results of the simulation using the scaled dataset across the three states, results of which were presented in the main manuscript text. Conversely, we find that cost estimates for hybrid power plants that provide 100 MW baseload generation based on just 8760h or one year of data, and therefore one year of simulation, are significantly lower than the cost estimates based on the full 20 year dataset and simulation period. 

\begin{table}[ht]
\caption{\textbf{\small Resource characteristics and optimal hybrid system that provides 100 MW baseload power every hour of the year in the state of Gujarat across different data sources for wind and solar generation. LCOE values are calculated assuming a 2.5\% cost of capital.}}
\begin{tabular}{|l|l|r|r|r|r|r|r|r|r|}
\hline
\textbf{State} & \textbf{\begin{tabular}[c]{@{}l@{}}Data \\ Source\end{tabular}} & \textbf{\begin{tabular}[c]{@{}l@{}}Wind \\ CF\end{tabular}} & \textbf{\begin{tabular}[c]{@{}l@{}}Solar \\ CF\end{tabular}} & \textbf{Years} & \textbf{\begin{tabular}[c]{@{}l@{}}Solar \\ (MW)\end{tabular}} & \textbf{\begin{tabular}[c]{@{}l@{}}Wind \\ (MW)\end{tabular}} & \textbf{\begin{tabular}[c]{@{}l@{}}Battery \\ (MWh)\end{tabular}} & \textbf{\begin{tabular}[c]{@{}l@{}}Solar (\%) in \\ wind-solar \\ mix\end{tabular}} & \textbf{\begin{tabular}[c]{@{}l@{}}LCOE \\ (Rs/kWh)\end{tabular}} \\
\hline
GJ & \begin{tabular}[c]{@{}l@{}}MERRA-2 \\ (Unscaled)\end{tabular} & 33\%                                               & 24\%                                                & 20    & 943        & 260       & 2038          & 78\%                         & 9.0           \\
\hline
GJ & \begin{tabular}[c]{@{}l@{}}MERRA-2 \\ (Scaled)\end{tabular}   & 23\%                                               & 20\%                                                & 20    & 1136       & 371       & 2038          & 75\%                         & 10.3          \\
\hline
GJ & Proprietary                                                   & 23\%                                               & 20\%                                                & 1     & 835        & 575       & 1577          & 59\%                         & 9.5           \\
\hline
GJ & NREL                                                          & 35\%                                               & 21\%                                                & 1     & 667        & 114       & 1689          & 85\%                         & 6.5 \\            
\hline         
\end{tabular}
\end{table}

Interestingly, we report that the hybrid systems designed to provide 100 MW baseload power that result from the optimization simulation using the proprietary one year average dataset, are in fact able to meet the 100 MW generation requirement in more than 98\% of hours across the 20 years if we run the full 20 year simulation using the scaled MERRA-2 dataset and fix the hybrid system capacities to the optimal system for the proprietary data. 

\begin{table}[ht]
\caption{\textbf{\small Resource characteristics and optimal hybrid system that provides 100 MW baseload power every hour of the year in the state of Tamil Nadu across different data sources for wind and solar generation. LCOE values are calculated assuming a 2.5\% cost of capital.}}
\begin{tabular}{|l|l|r|r|r|r|r|r|r|r|}
\hline
\textbf{State} & \textbf{\begin{tabular}[c]{@{}l@{}}Data \\ Source\end{tabular}} & \textbf{\begin{tabular}[c]{@{}l@{}}Wind \\ CF\end{tabular}} & \textbf{\begin{tabular}[c]{@{}l@{}}Solar \\ CF\end{tabular}} & \textbf{Years} & \textbf{\begin{tabular}[c]{@{}l@{}}Solar \\ (MW)\end{tabular}} & \textbf{\begin{tabular}[c]{@{}l@{}}Wind \\ (MW)\end{tabular}} & \textbf{\begin{tabular}[c]{@{}l@{}}Battery \\ (MWh)\end{tabular}} & \textbf{\begin{tabular}[c]{@{}l@{}}Solar (\%) in \\ wind-solar \\ mix\end{tabular}} & \textbf{\begin{tabular}[c]{@{}l@{}}LCOE \\ (Rs/kWh)\end{tabular}} \\
\hline
TN    & \begin{tabular}[c]{@{}l@{}}MERRA-2 \\ (Unscaled)\end{tabular} & 25\%    & 18\%     & 20    & 1679       & 317       & 1933          & 84\%                         & 11.7          \\
\hline
TN    & \begin{tabular}[c]{@{}l@{}}MERRA-2 \\ (Scaled)\end{tabular}   & 19\%    & 20\%     & 20    & 1550       & 109       & 2606          & 93\%                         & 11.5          \\
\hline
TN    & Proprietary                                                   & 19\%    & 20\%     & 1     & 1085       & 31        & 1629          & 97\%                         & 7.4           \\
\hline
TN    & NREL                                                          & 42\%    & 20\%     & 1     & 1219       & 0         & 2520          & 100\%                        & 9.5      \\        
\hline         
\end{tabular}
\end{table}

For instance in the case of Karnataka (Table S9), the optimal system based on the average one year of proprietary data has a levelized cost of \rupee 6.6/kWh assuming a 2.5\% cost of capital, which is more than 40\% cheaper than the system based on the scaled MERRA-2 dataset. If we run the optimization model with the scaled MERRA-2 dataset as input, and a fixed system design with the objective of maximizing the availability, i.e. the number of hours that the system meets the 100 MW generation requirement, we find that the smaller system based on the proprietary data (745 MW solar, 112 MW wind and 1593 MWh battery) can meet the generation requirement 98.7\% of hours across 20 years. 

This goes to show that just a small portion of hours in the 20 year dataset are significantly influencing the design of the system and driving up costs. These results are entirely consistent with the literature on the importance of modelling variable renewable resources over longer timescales due to tail events that impact wind and solar output. For example, see Sepulveda et al (2021)\cite{sepulveda2021design} and Dowling et al (2020)\cite{dowling2020role}.

The same holds true for Tamil Nadu (Table S11). We find that the smaller system based on the proprietary data in Tamil Nadu (1085 MW solar, 31 MW wind and 1629 MWh battery) can meet the generation requirement in 98.4\% of hours across twenty years. In Gujarat (Table S10) there is no major difference in costs or sizing between the optimal hybrid system based on the proprietary data and the MERRA-2 data. As such, we indeed find that the proprietary data based system (835 MW solar, 575 MW wind and 1577 MWh battery) can actually meet the 100 MW generation requirement in 99.9\% of hours across twenty years. 

The differences between the model results based on the twenty year wind and solar dataset and the one year data for Karnataka and Tamil Nadu indicate that our twenty year analysis and data are more likely to account for rare events (for e.g. storms) that might lead to solar and wind outages. This significant inter-annual variability is clearly better captured in the twenty year simulation while missing in the one year simulation. Clearly, when designing a capital asset such as a hybrid power plant that will operate for twenty years and has stringent generation requirements, a consideration of the frequency of rare events that can impact wind and solar output will be critical. In our case, the variability of wind and solar affects the sizing of the battery system, and therefore heavily influences estimates of levelized costs.

We also find that the levelized cost estimates for systems based on the one year datasets are in the range of \rupee 7-9 /kWh which is exactly the same range as the levelized costs of 99\% availability systems based on the 20 year MERRA-2 scaled data, which were shown in Figure 5 in the main manuscript text and Tables S1-S3 in Note 1. Again this shows that modelling based on one year of data misses out on rare prolonged outages of wind and solar, which are better captured in data spread across longer time periods. Even if these events are only present in 1\% of hours across twenty years, they nonetheless lead to large increases in costs.

Overall, across the different data inputs for wind and solar generation at our site locations in the three states, we find that our results, i.e. suitability of solar compared to wind for pairing with lithium-ion battery storage and the high cost of hybrid systems today compared to existing coal plants in India, are robust. Put simply, the main policy conclusions of our analysis are unchanged despite using differing data sources.

In particular, we stress that despite much higher CFs for wind in certain data (for e.g. NREL wind CFs for Tamil Nadu and Gujarat shown in Tables S10, S11), optimal hybrid systems based on those data continue to be solar dominated. This is due to the fact that wind in India is extremely seasonal with extremely high output during the monsoon period followed by very limited production in the rest of the year. As a result, it is unsuitable for pairing with storage to provide firm generation to the grid throughout the year. We further explore the characteristics of the wind and solar resource in India in Supplementary Note 4.

\clearpage
\section{Supplementary Note 4: Exploring wind and solar resources in India}

Wind energy is highly seasonal with high output during the monsoon months and limited production rest of the year. To verify this in our different data sources, we offer an exploration of the data below. We define seasonality here as the amount of production in hours 4000-6000 (roughly mid June - mid September which is monsoon period in India) compared to the amount of production the rest of the year. For a resource that is distributed equally throughout the year, the ratio should simply be 2000/6760 = 0.30. The greater the divergence from 0.3, the more the seasonality in either direction. We find that for wind, as shown in Table S12 below, the seasonality ratio is mostly greater than 0.7, and sometimes even higher than 1, which demonstrates how high wind power generation is during the monsoon season compared to the rest of the year. Solar on the other hand can be clearly seen to be not nearly as seasonal with the ratio close to 0.3 but lower, indicating that solar production naturally drops during the monsoon months but it is not as drastically affected as wind production, which rises significantly compared to other months of the year. 

This high seasonality for wind power in India makes it unsuitable for pairing with short duration storage in hybrid systems that must operate 100\% of the year providing baseload or flexible generation. Aside from the issue of seasonality, tables S9-S11 in the previous section also showed the differing capacity factors for wind and solar in the three states across different data sources. In particular, wind capacity factor varies significantly across different data sources, with solar more consistent. 

\begin{table}[!ht]
\begin{center}
\caption{\textbf{\small : Seasonality and capacity factor of wind and solar energy across different data sources for the proprietary site locations in the three states. Wind turbine and solar panel configurations were kept constant within each state across all data sources to enable direct comparison.}}
\begin{tabular}{|r|l|l|l|r|r|r|r|}
\hline
\multirow{2}{*}{\textbf{\#}} & \multirow{2}{*}{\textbf{State}} & \multirow{2}{*}{\textbf{Wind \& Solar Dataset}} & \multirow{2}{*}{\textbf{\begin{tabular}[c]{@{}l@{}}Data \\ duration\end{tabular}}} & \multirow{2}{*}{\textbf{\begin{tabular}[c]{@{}l@{}}Wind \\ Seasonality\end{tabular}}} & \multirow{2}{*}{\textbf{\begin{tabular}[c]{@{}l@{}}Solar \\ Seasonality\end{tabular}}} & \multirow{2}{*}{\textbf{\begin{tabular}[c]{@{}l@{}}Solar \\ CF\end{tabular}}} & \multirow{2}{*}{\textbf{\begin{tabular}[c]{@{}l@{}}Wind \\ CF\end{tabular}}} \\
                             &                                 &                                                 &                                                                                    &                                                                                       &                                                                                        &                                                                               &                                                                              \\
\hline                             
1                            & \multirow{4}{*}{Karnataka}      & Proprietary                                     & 1 year                                                                             & 0.588                                                                                 & 0.237                                                                                  & 19\%                                                                          & 22\%                                                                         \\
2                            &                                 & MERRA-2 2000-2019                               & 20 years                                                                           & 0.882                                                                                 & 0.206                                                                                  & 22\%                                                                          & 22\%                                                                         \\
3                            &                                 & NREL 2014                                       & 1 year                                                                             & 0.849                                                                                 & 0.232                                                                                  & 21\%                                                                          & 25\%                                                                         \\
4                            &                                 & MERRA-2 2014                                    & 1 year                                                                             & 0.989                                                                                 & 0.200                                                                                  & 22\%                                                                          & 22\%                                                                         \\
\hline  
5                            & \multirow{4}{*}{Gujarat}        & Proprietary                                     & 1 year                                                                             & 0.467                                                                                 & 0.224                                                                                  & 20\%                                                                          & 23\%                                                                         \\
6                            &                                 & MERRA-2 2000-2019                               & 20 years                                                                           & 0.478                                                                                 & 0.221                                                                                  & 24\%                                                                          & 33\%                                                                         \\
7                            &                                 & NREL 2014                                       & 1 year                                                                             & 0.583                                                                                 & 0.248                                                                                  & 21\%                                                                          & 35\%                                                                         \\
8                            &                                 & MERRA-2 2014                                    & 1 year                                                                             & 0.538                                                                                 & 0.238                                                                                  & 24\%                                                                          & 32\%                                                                         \\
\hline  
9                            & \multirow{4}{*}{Tamil Nadu}     & Proprietary                                     & 1 year                                                                             & 1.228                                                                                 & 0.277                                                                                  & 20\%                                                                          & 19\%                                                                         \\
10                           &                                 & MERRA-2 2000-2019                               & 20 years                                                                           & 0.739                                                                                 & 0.222                                                                                  & 18\%                                                                          & 25\%                                                                         \\
11                           &                                 & NREL 2014                                       & 1 year                                                                             & 0.886                                                                                 & 0.274                                                                                  & 20\%                                                                          & 42\%                                                                         \\
12                           &                                 & MERRA-2 2014                                    & 1 year                                                                             & 0.827                                                                                 & 0.227                                                                                  & 18\%                                                                          & 26\%      \\
\hline  
\end{tabular}
\end{center}
\vspace{-0.2cm}
\end{table}

Much of the previous energy modelling literature on India has been based on the NREL dataset\cite{nrel}. While this dataset captures the high seasonality of wind as shown in Table S12, consistent with other data sources, capacity factor ratings for wind are higher than in other data, particularly for the state of Tamil Nadu where we see a large difference compared to other data sources, including the MERRA-2 data source for the same year as the NREL dataset (2014). 

We find that the NREL dataset's wind capacity factor of $>$40\% for Tamil Nadu is a significant outlier compared to the observed data of capacity factors for wind energy in India, which are usually in the range of 20-30\%. In Table S13, we compile observed production data from wind energy across all of India between November 2020-October 2021. Table S13 shows the monthly capacity factor for wind in India as well as the 12 month average. Firstly, we can see that all India wind data in Table S13 also reflects the seasonality of wind, the months of June-September have significantly higher generation than the rest of the year. Second, we see that the 12 month average capacity factor is just 20\%. Note that wind power in India is concentrated in just 7 states, including our selected states of Karnataka, Gujarat, and Tamil Nadu. As such, these states are well represented in these data.

Finally, in Table S14 we show the year on year CFs for wind and solar in the state of Karnataka from our 20 year MERRA-2 based dataset, to reflect the high inter-annual variability for wind compared to solar power. For example, in 2011, wind CF is 26\%, compared to just 19\% a year previously in 2010. Solar CFs only vary between 21-23\% over the entire period.

Overall, our exploration of wind and solar resources in India results in a few key findings. Tables S12, S13, and S14 show that solar resources in India have limited variation both spatially (across different states) and temporally (across years or seasons). On the other hand, wind energy can have significant variation from year to year (Table S14), across different regions (Table S12), and has high seasonality with most of the annual generation occurring during the monsoon period from June to September (Table S12, S13).

\begin{table}[!hbt]
\caption{\textbf{\small All India observed wind generation statistics between November 2020 - October 2021. Data source: Central Electricity Authority\cite{cea2021}.}}
\vspace{-0.9cm}
\begin{center}
\begin{tabular}{|l|r|r|r|}
\multicolumn{4}{c}{\textbf{ALL INDIA WIND   2020-2021}}  
\\
\hline
\textbf{Month} & \textbf{Installed Capacity (GW)}                                                            & \textbf{Total Generation (GWh)}                                                       & \textbf{Calculated Monthly CF}                                   \\
\hline
November       & 38.43                                                                                       & 3305                                                                                  & 12\%                                                             \\
\hline
December       & 38.62                                                                                       & 3815                                                                                  & 13\%                                                             \\
\hline
January        & 38.68                                                                                       & 3562                                                                                  & 12\%                                                             \\
\hline
February       & 38.79                                                                                       & 2873                                                                                  & 11\%                                                             \\
\hline
March          & 39.25                                                                                       & 3532                                                                                  & 12\%                                                             \\
\hline
April          & 39.41                                                                                       & 3734                                                                                  & 13\%                                                             \\
\hline
May            & 39.44                                                                                       & 7084                                                                                  & 24\%                                                             \\
\hline
June           & 39.49                                                                                       & 9725                                                                                  & 34\%                                                             \\
\hline
July           & 39.59                                                                                       & 11421                                                                                 & 39\%                                                             \\
\hline
August         & 39.69                                                                                       & 9349                                                                                  & 32\%                                                             \\
\hline
September      & 39.87                                                                                       & 6981                                                                                  & 24\%                                                             \\
\hline
October        & 39.99                                                                                       & 2978                                                                                  & 10\%                                                             \\
\hline
\textbf{12 months}      & \begin{tabular}[r]{@{}r@{}}\textbf{39.27} \\   (12 months average)\end{tabular} & \begin{tabular}[r]{@{}r@{}}\textbf{68359} \\ (total 12 months   generation)\end{tabular} & \begin{tabular}[r]{@{}r@{}}\textbf{20\%} \\    (12 month)\end{tabular}\\
\hline
\end{tabular}
\end{center}
\end{table}

\clearpage
\begin{longtable}[!tbp]{|l|r|r|r|r|r|}
\caption{\textbf{\small Seasonality and capacity factor of wind and solar energy across different years for the site location in Karnataka}}\\
\hline
\textbf{\#} & \textbf{\begin{tabular}[c]{@{}l@{}}Wind   \& Solar \\ Data   Source\end{tabular}} & \textbf{\begin{tabular}[c]{@{}l@{}}Wind   \\ Seasonality\end{tabular}} & \textbf{\begin{tabular}[c]{@{}l@{}}Solar   \\ Seasonality\end{tabular}} & \textbf{Solar CF} & \textbf{Wind CF} \\
\hline
1           & MERRA-2 2000-2019                                                             & 0.882                                                                  & 0.206                                                                   & 22\%              & 22\%             \\
\hline
2           & MERRA-2 2000                                                                            & 0.744                                                                  & 0.204                                                                   & 23\%              & 23\%             \\
\hline
3           & MERRA-2 2001                                                                            & 0.880                                                                  & 0.226                                                                   & 22\%              & 24\%             \\
\hline
4           & MERRA-2 2002                                                                            & 0.983                                                                  & 0.228                                                                   & 23\%              & 23\%             \\
\hline
5           & MERRA-2 2003                                                                            & 0.859                                                                  & 0.182                                                                   & 23\%              & 22\%             \\
\hline
6           & MERRA-2 2004                                                                            & 0.708                                                                  & 0.218                                                                   & 23\%              & 24\%             \\
\hline
7           & MERRA-2 2005                                                                            & 1.015                                                                  & 0.192                                                                   & 21\%              & 21\%             \\
\hline
8          & MERRA-2 2006                                                                            & 0.795                                                                  & 0.213                                                                   & 23\%              & 26\%             \\
\hline
9          & MERRA-2 2007                                                                            & 0.974                                                                  & 0.186                                                                   & 21\%              & 22\%             \\
\hline
10          & MERRA-2 2008                                                                            & 0.623                                                                  & 0.197                                                                   & 22\%              & 22\%             \\
\hline
11          & MERRA-2 2009                                                                            & 0.941                                                                  & 0.222                                                                   & 21\%              & 22\%             \\
\hline
12        & MERRA-2 2010                                                                            & 0.972                                                                  & 0.210                                                                   & 20\%              & 19\%             \\
\hline
13          & MERRA-2 2011                                                                            & 0.813                                                                  & 0.196                                                                   & 23\%              & 26\%             \\
\hline
14         & MERRA-2 2012                                                                            & 0.922                                                                  & 0.198                                                                   & 22\%              & 22\%             \\
\hline
15          & MERRA-2 2013                                                                            & 0.888                                                                  & 0.201                                                                   & 22\%              & 25\%             \\
\hline
16          & MERRA-2 2014                                                                            & 0.989                                                                  & 0.200                                                                   & 22\%              & 22\%             \\
\hline
17          & MERRA-2 2015                                                                            & 0.949                                                                  & 0.236                                                                   & 22\%              & 20\%             \\
\hline
18          & MERRA-2 2016                                                                            & 0.924                                                                  & 0.209                                                                   & 23\%              & 20\%             \\
\hline
19          & MERRA-2 2017                                                                            & 0.736                                                                  & 0.207                                                                   & 22\%              & 20\%             \\
\hline
20          & MERRA-2 2018                                                                            & 1.173                                                                  & 0.213                                                                   & 22\%              & 22\%             \\
\hline
21          & MERRA-2 2019                                                                            & 0.997                                                                  & 0.211                                                                   & 22\%              & 22\%  
\hline
\end{longtable}

\clearpage

\section{Supplementary Note 5: Simulation based on Karnataka's observed net load data}
We collected state wide hourly net load data (state wide hourly demand minus state wide hourly production from intermittent renewable energy) for Karnataka for the year 2018 from the Karnataka Power Transmission Corporation Limited (KPTCL). Data were obtained under non disclosure agreement from KPTCL. While all results of our optimization model presented in the main paper text and in the SI thus far are based on stylized coal profiles shown in Figure 1(a,b) in the main manuscript text, here we present an additional analysis using observed normalized (to 100 MW) hourly net load data as the target coal generation profile in our optimization model. We run this additional robustness check to verify that our stylized profiles are not driving results.

\begin{figure}[!h]
    \centering
    \includegraphics[width=9cm, height = 6.7cm]{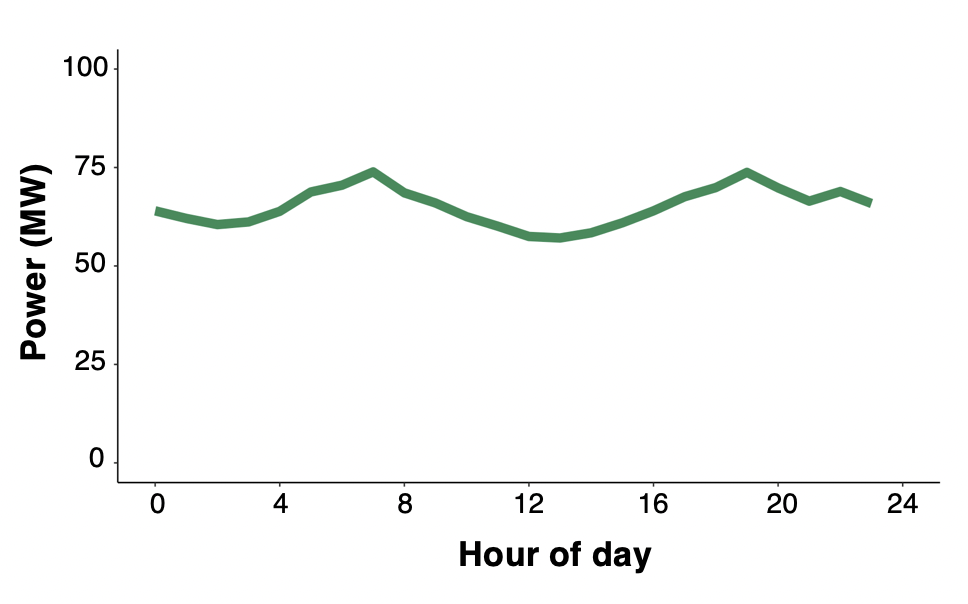}
    \caption{Daily average of normalized net load for Karnataka for the year 2018}
    \label{fig10}
\end{figure}

First, in Figure S\ref{fig10} above we show the daily average of the proprietary normalized hourly annual net load for Karnataka. Note that the daily average here is just for illustration purposes, the optimization model was run using the entire 8760h proprietary hourly net load data which was repeated for 20 years, i.e. in the absence of observed 20 year net load data, we use the one year net load and assume that repeats each year for 20 years. Our data inputs for wind and solar resource are once again the scaled MERRA-2 dataset as in the main manuscript text. 

In Table S15 we show the levelized costs and design of a hybrid system that meets the Karnataka net load profile of generation for all 175200 hours across twenty years. These levelized costs are assuming a 2.5\% cost of capital. We compare these costs to the results (also shown in Figure 2 and Table 2 in the main manuscript text) for the stylized flexible profile. Clearly, the hybrid system and levelized costs for this observed net load profile are not too dissimilar, indicating that the stylized profile we use does not distort our results.

\begin{table}[ht]
    \vspace{0.5cm}
  \begin{center}
    \caption{\textbf{\small Levelized costs and capacity mixes of the least cost hybrid power plants that combine wind power, solar PV, and 4-hour duration Li-ion battery storage to provide a flexible generation profile (stylized and observed). Reported LCOE values are for a 2.5\% WACC.}}\label{table7}
    \vspace{0.5cm}
    \begin{tabular}{l|r|r|r|r} 
      \textbf{Profile} & \textbf{LCOE (\rupee/kWh)} & \textbf{Solar (MW)} & \textbf{Wind (MW)} & \textbf{Battery Storage (MWh)}\\
      \hline
      Stylized & 13.8 & 1582 & 0 & 2553\\
      KPTCL Net load & 12.3 & 1006 & 94 & 1897\\
    \hline
    \end{tabular}
  \end{center}
\end{table}

\clearpage
\section*{References}
\bibliography{supplement_cites.bib}